\documentclass[useAMS,usenatbib]{mn2e}
\usepackage{graphicx}
\usepackage{times}
\usepackage{amssymb}
\usepackage{natbib}
\def\gsim { \lower .75ex \hbox{$\sim$} \llap{\raise .27ex \hbox{$>$}} }
\def\lsim { \lower .75ex \hbox{$\sim$} \llap{\raise .27ex \hbox{$<$}} }

\begin{document}

\title[Satellites and Fossil Groups in the Millennium Simulation]
{Satellite Galaxies and Fossil Groups in the Millennium Simulation}

\author[Sales et al.]{
\parbox[t]{\textwidth}{
Laura V. Sales$^{1,2}$,
Julio F. Navarro$^{3,4}$\thanks{Fellow of the Canadian Institute for Advanced Research},
Diego G. Lambas$^{1,2}$,
Simon D. M. White$^{4}$ \&
Darren J. Croton$^{5}$
}
\vspace*{6pt} \\
\\
$^{1}$ Observatorio Astron\'omico, Universidad Nacional de C\'ordoba, Laprida
854, 5000 C\'ordoba, Argentina.
\\
$^{2}$ Instituto de Astronom\'{\i}a Te\'orica y Experimental, Conicet, Argentina.\\
$^3$Department of Physics and Astronomy, University of Victoria, B.C., Canada.\\
$^4$Max-Planck-Institut f\"ur Astrophysik, D-85740 Garching, Germany.\\
$^5$Department of Astronomy, University of California, Berkeley, CA, 94720, USA.}

\maketitle

\begin{abstract}
We use a semianalytic galaxy catalogue constructed from the {\it
Millennium Simulation} ({\it MS}) to study the satellites of isolated
galaxies in the $\Lambda$CDM cosmogony. The large volume surveyed by
the {\it MS} ($500^3 \, h^{-3}{\rm Mpc}^3$), together with its
unprecedented numerical resolution, enable the compilation of a large
sample of $\sim 80,000$ bright ($M_r<-20.5$) primaries, surrounded by
$\sim 178,000$ satellites down to the faint magnitude limit
($M_r=-17$) of our catalogue. This sample allows the characterization,
with minimal statistical uncertainty, of the dynamical properties of
satellite/primary galaxy systems in a $\Lambda$CDM universe. The
details of this characterization are sensitive to the details of the
modeling, such as its assumptions on galaxy merging and dynamical
friction timescales, but many of its general predictions should be
applicable to hierarchical formation models such as $\Lambda$CDM. We
find that, overall, the satellite population traces the dark matter
rather well: its spatial distribution and kinematics may be
approximated by an NFW profile with a mildly anisotropic velocity
distribution.  Their spatial distribution is also mildly anisotropic,
with a well-defined ``anti-Holmberg'' effect that reflects the
misalignment between the major axis and angular momentum of the host
halo. Our analysis also highlights a number of difficulties afflicting
studies that rely on satellite velocities to constrain the primary
halo mass. These arise from variations in the star formation
efficiency and assembly history of isolated galaxies, which result in
a scatter of up to $\sim 2$ decades in halo mass at fixed primary
luminosity. Our isolation criterion (primaries may only have
companions at least 2 mag fainter within $1 \, h^{-1}$ Mpc)
contributes somewhat to the scatter, since it picks not only galaxies
in sparse environments, but also a number of primaries at the centre
of ``fossil'' groups. We find that the abundance and luminosity
function of these unusual systems are in reasonable agreement with the
few available observational constraints. Much tighter halo
mass-luminosity relations are found when splitting the sample by
colour: red primaries inhabit halos more than twice as massive as
those surrounding blue primaries, a difference that vanishes, however,
when considering stellar mass instead of luminosity. The large scatter
in the halo mass-luminosity relation hinders the interpretation of the
velocity dispersion of satellites stacked according to the luminosity
of the primary. We find $L\propto \sigma^3$ (the natural scaling
expected for $\Lambda$CDM) for truly-isolated primaries, i.e., systems
where the central galaxy contributes more than $85\%$ of the total
luminosity within its virial radius. Less strict primary selection,
however, leads to substantial modification of the scaling relation:
blindly stacking satellites of all primaries results in a much
shallower $L$-$\sigma$ relation that is only poorly approximated by a
power law.
\end{abstract}

\date{}


\section{INTRODUCTION}
\label{sec:intro}

Satellite galaxies may be thought of as the visible fossil relics of
hierarchical galaxy formation, where the mass of a galaxy is
envisioned to be assembled in a sequence of accretion events. As
surviving witnesses of the accretion process, satellites bear an
invaluable record of the assembly history of the primary galaxy they
orbit, and provide at the same time prime information about the mass
and extent of the dark matter halo they inhabit.

Satellites are a particularly valuable tool for studying the outer
regions of dark matter halos, where few other tracers exist that can
provide effective constraints. This is especially true in the case of
the Local Group, where the dynamics of the outer satellites has played
an important role in mass estimates of the Milky Way and M31
\citep{littleandtremaine87,zaritsky89,kochanek96,wilkinson99,
evans00,battaglia05}. Local Group satellites can be found even if they
are extremely faint; Draco, for example, is $\sim 8 \times 10^{4}$
times fainter than its primary, the Milky Way, and some of the new
satellites discovered over the past few years have luminosities
rivaling that of ordinary star clusters
\citep{zucker04,zucker06,willman05b,martin06,belokurov06,
belokurov07,irwin07,majewski07,ibata07,chapman07}.

On the other hand, extragalactic satellite studies have been
traditionally limited by the scarcity of satellite/primary systems
easily accessible to observation. Partly as a result of the strict
isolation criteria that are imposed on the primaries in order to
minimize interlopers and to avoid complications that may arise from
having more than one dominant object, it is rare for primaries to have
more than a few satellites bright enough for spectroscopic
confirmation and follow up.

The scarcity of satellites surrounding one given primary has prompted
the adoption of stacking techniques in order to overcome small number
statistics. For example, the satellites of all primaries of given
luminosity, $L$, might be combined to yield estimates of the {\it
average}, rather than individual, halo properties as a function of
$L$.  These techniques are clearly vulnerable to the presence of
luminosity-dependent biases in the satellite distribution and of
systematic trends between halo mass and primary luminosity that may be
difficult to detect and evaluate in observational datasets.

Such techniques, nevertheless, seem appropriate to address a number of
important issues in galaxy formation studies.  How does the average
dark halo mass depend on the luminosity of the primary? What is the
mass profile of the dark halo and how does it vary with luminosity?
What is the three-dimensional shape of the dark halo and how does it
relate to the primary and to the spatial distribution of satellites?
These are some of the questions traditionally addressed by extragalactic
satellite studies, and there is a rich literature documenting prior
attempts at exploiting the dynamics of satellites to constrain the
mass and extent of dark matter halos.

The pioneering work of \citet{zaritsky93,zaritsky97a}, for example,
confirmed the presence of massive halos around isolated spirals, 
but also hinted at a few odd results that are difficult to 
reconcile with current galaxy formation models. For example, these 
authors noted that satellite velocities seemed to be independent 
of the luminosity of the primary, contrary to what might be naively
expected from the Tully-Fisher relation. They also remarked that 
satellites seemed to be distributed anisotropically around the 
primary, in agreement with the early suggestion of \citet{holmberg69}.

These issues have been revisited using the much larger datasets
compiled by the 2dfGRS \citep{colless01} and the SDSS
\citep{york00,strauss02} surveys, and broad consensus seems to be
gradually emerging.  In particular, \citet{mckay02} and
\citet{prada03} find that a well-defined trend between satellite
velocities and primary luminosity {\it does} appear when considering
samples substantially larger than the ones considered by Zaritsky et
al. Better statistics have also clarified the anisotropic distribution
of satellites around spirals, which, contrary to Holmberg's
suggestion, apparently tend to avoid the rotation axis of the disk
\citep{brainerd05,azzaro06,yang06,agustsson07}.  Finally, the
availability of larger samples have also enabled studies of the
satellite velocity dispersion profile, which may be used to probe the
outer dark mass distribution and to compare it with cosmological
N-body simulations (see \citealt{prada03}, also the review of
\citealt{brainerd04b}, and van den Bosch et al. 2005a,b).
\nocite{vandenbosch05a,vandenbosch05b}

The success of these studies depends crucially on identifying and
understanding possibly subtle biases between dark matter and the
satellite population. This is best accomplished through direct
numerical simulations, where full 3D dynamical information is
available and from which mock observational datasets may be created to
assess the ability of analysis techniques to identify and correct for
such biases. Progress, however, has been slow, mainly because
simulations with enough dynamic range to resolve simultaneously a
galaxy and its satellites have only recently become possible
\citep{klypin99a,moore99,kravtsov04,gao04b,diemand04,maccio06,
diemand07,libeskind07,sales07a,sales07b}. Further, these simulations
have typically followed small volumes (single halos, in many cases)
and many of them have focussed on the dark matter component only,
hindering direct comparisons between theory and observation. 

We investigate these issues here using the galaxy catalogue created
by \citet{croton06} from the {\it Millennium Simulation} 
(\citealt{springel05}, hereafter {\it MS} for short).  The catalogue is the
result of a sophisticated semianalytic galaxy formation model applied
to an N-body simulation of unprecedented dynamic range. This model has
been calibrated to reproduce many of the large-scale properties of the
observed galaxy population, such as the luminosity and correlation
functions, as well as their color dependence. However, no specific
information about satellite systems has been taken into account and,
therefore, the results we present here may be considered true
theoretical ``predictions'' that can be contrasted fruitfully with
observation.

Our principal aim is to provide a detailed characterization of the 3D
properties of the satellite population of bright, isolated galaxies in
the $\Lambda$CDM cosmogony. This characterization is intended (i) to
guide the interpretation of observational datasets, (ii) to improve
the identification of primary/satellite systems so as to minimize the
contamination by interlopers, and (iii) to provide a framework within
which the questions posed above may be profitably addressed.  The
sheer size of the catalogue, which lists $\sim 10^7$ galaxies brighter than
$M_r=-17$ in the $500^3 h^{-3}$ Mpc$^3$ volume of the {\it MS}
{\footnote{We express the present-day value of the Hubble constant as
$H_0=100\, h$ km/s/Mpc.  Throughout this paper, masses and
luminosities will assume $h=0.73$ unless otherwise specified.}, ensures
that our results have negligible statistical uncertainty.

Our plan for the paper is as follows. After a brief description of the
{\it MS} (\S~\ref{ssec:MS}) and of the semianalytic galaxy formation
model (\S~\ref{ssec:sam}), we discuss our selection criteria for
primaries and for satellites in \S~\ref{ssec:gxcat}. Our main results
are presented and discussed in \S~\ref{sec:results}. We begin in
\S~\ref{ssec:lmc} with a discussion of the relation between halo mass,
primary luminosity and color. Noting that our isolation criteria for
primaries allows for the inclusion of ``fossil groups'' in our sample,
we compare their abundance and luminosity function to observations in
\S~\ref{ssec:fossil}. The satellites' spatial distribution is
discussed next (\S~\ref{ssec:satprof}), together with their kinematics
(\S~\ref{ssec:satvel}). We end the discussion of our results by
analyzing the relation between primary luminosity and satellite
velocity dispersion (\S~\ref{ssec:lsg}), as well as the presence of
anisotropies (\S~\ref{ssec:holm}) in the satellite spatial
distribution. We conclude with a brief summary in \S~\ref{sec:conc}.

\begin{figure*}
\begin{center}
\includegraphics[width=\linewidth,clip]{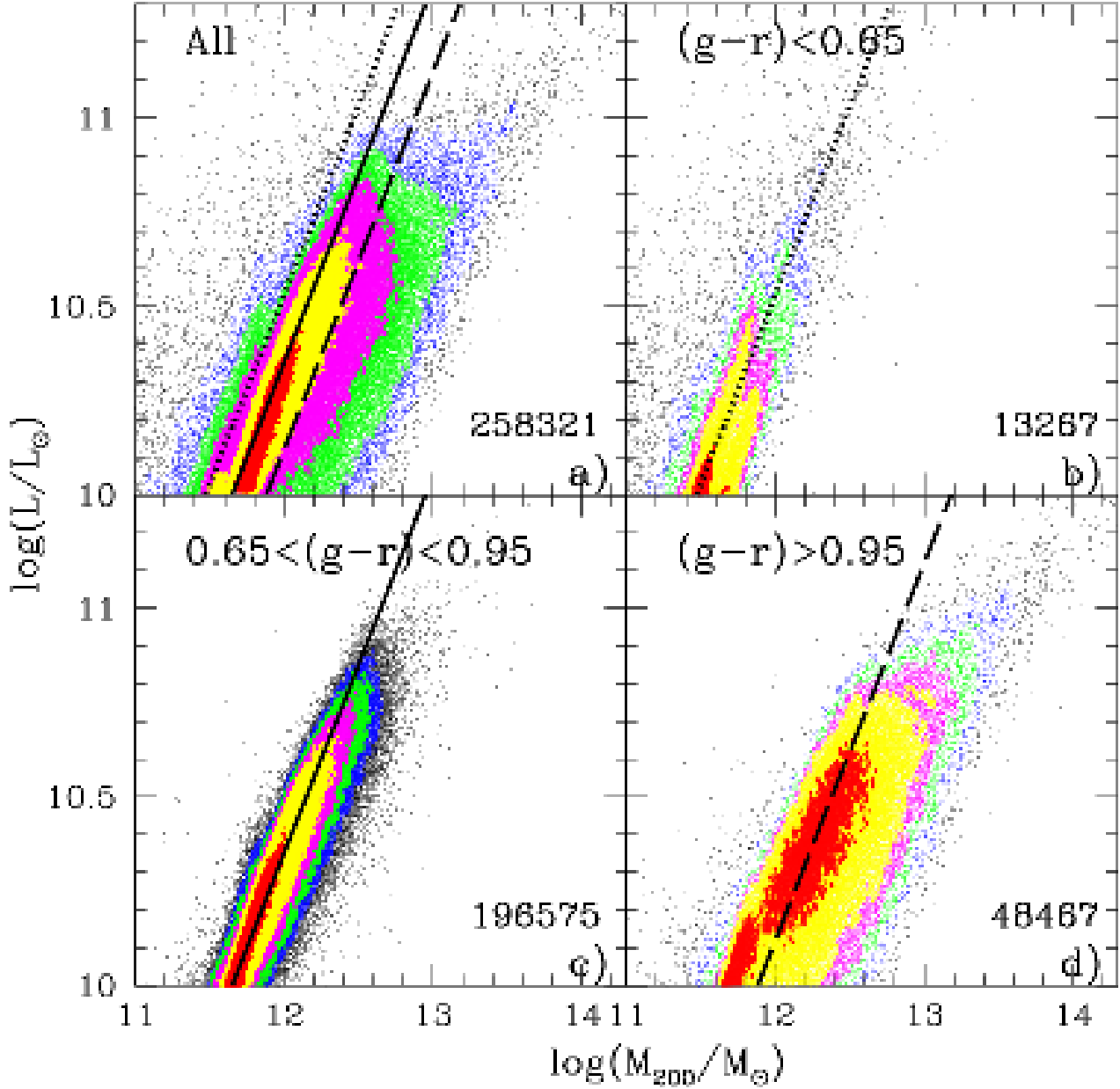}
\end{center}
\caption{Virial mass vs r-band luminosity relation for primary
galaxies.  Panel $(a)$ shows data for all primaries, other panels are
subsamples after applying a color cut (see labels). The total number
of primaries plotted in each panel is quoted.  Colors indicate the
density of primaries at various locations in each panel, in
logarithmic units; each color step corresponds to a variation of about
10 in number. The same applies to other panels, although the color
coding has been renormalized in each case to make use of the whole
color palette.  Dotted, solid and dashed lines indicate loci of
constant virial mass-to-light ratio, and are chosen to ease the
comparison between panels and to outline the presence of three
``sequences'' in the data (see text for further discussion).}
\label{figs:lmc}
\end{figure*}

\begin{figure*}
\begin{center}
\includegraphics[width=0.475\linewidth,clip]{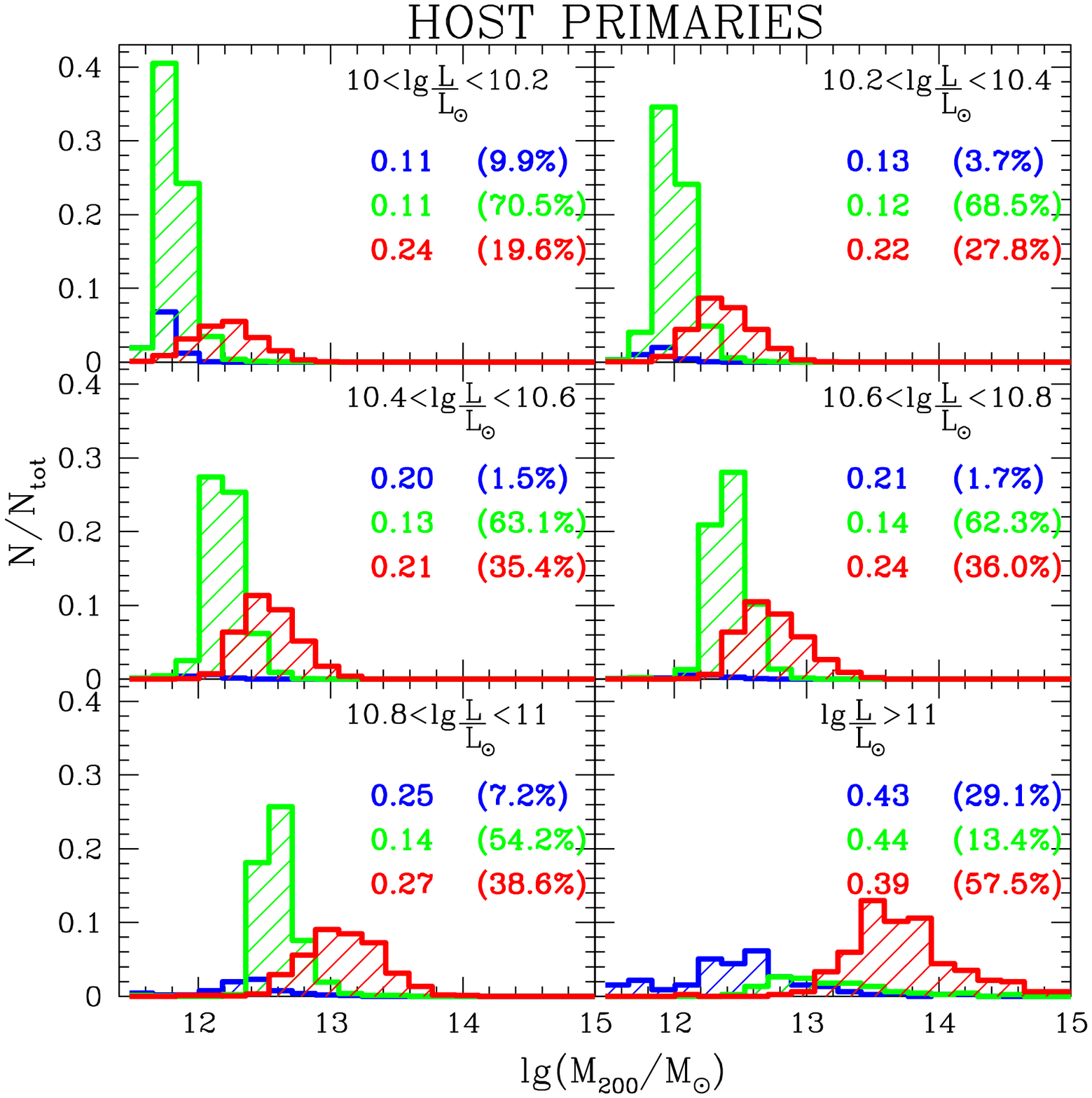}
\hspace{0.4cm}
\includegraphics[width=0.475\linewidth,clip]{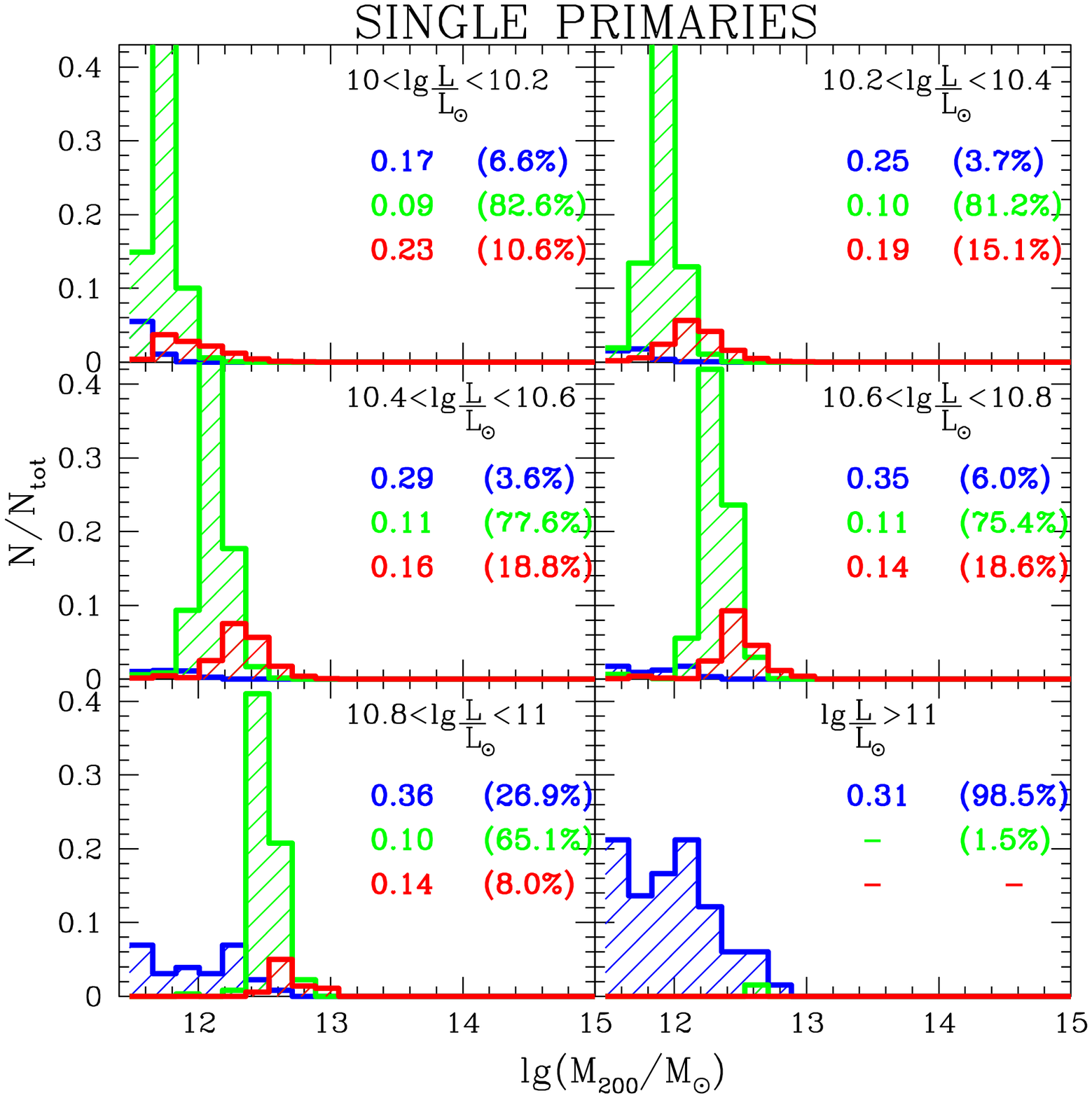}
\caption{Virial mass distribution in luminosity bins for {\it host}
(left panel) and {\it single} (right panel) primaries, split according
to color (see Figure~\ref{figs:lmc}). Histograms have been normalized
to the total number of galaxies in each luminosity bin. For {\it
hosts} we have: $(12,747)$, $(23,585)$, $(26,039)$, $(13,817)$, $(2,436)$ and
$(454)$ galaxies in each of the $6$ bins of increasing $L$,
respectively. For {\it single} primaries, the corresponding numbers
are : $(91,060)$, $(58,364)$, $(23,893)$, $(4,742)$, $(361)$ and $(66)$,
respectively.  Percentages quoted in the right column of each panel indicate 
the fraction of primaries in each color sequence for a given luminosity bin.
Left columns show the rms values $\sigma(\rm log(M_{200}/M_\odot))$ 
corresponding to each subsample.}
\label{figs:lmcdist}
\end{center}
\end{figure*}

\section{The Catalogue}

\subsection{The Millennium Simulation}
\label{ssec:MS}

The {\it{Millennium Simulation}} ({\it MS}), one of the projects of
the Virgo Consortium \footnote{http://www.virgo.dur.ac.uk}, is a
cosmological N-body simulation of the $\Lambda$CDM universe that
follows the evolution of more than 10 billion particles in a box of
500 $h^{-1}$ Mpc (comoving) on a side. The particle mass is $8.6
\times 10^9 \, h^{-1} M_\odot$, and particle-particle gravitational
interactions are softened on scales smaller than 5 $h^{-1}$ kpc.  The
simulation adopts parameters consistent with the WMAP1 results
\citep{spergel03,seljak05}: $\Omega_{\rm m}=0.25$,
$\Omega_{\Lambda}=0.75$, $h=0.73$, $n=1$, and $\sigma_8=0.9$ (for
details see Springel et al. 2005). Data from the simulation output
at 64 times spaced logarithmically in expansion factor before $z=1$,
and at approximately $200$ Myr intervals thereafter. These data are
used to build merger trees which encode the assembly history of each
halo and its resolved substructure. These form the basis for the
galaxy formation modeling. Complete data for these halo merging trees
as well as for several galaxy formation models can be found at {\small \tt
http://www.mpa-garching.mpg.de/millennium}. The $z=0$ data from the
Croton et al (2006) model used here can be found at 
{\small \tt http://www.mpa-garching.mpg.de/galform/agnpaper/}.

\subsection{The Semianalytic Model}
\label{ssec:sam}

The information contained in the {\it MS} is harvested using
semianalytic galaxy formation models that follow, with simple, but
physically motivated, laws the formation of galaxies within the
evolving dark matter halos. The actual prescriptions used derive from
the work of \citet{kauffmann99}, \citet{springel01} and
\citet{delucia04b}, but have been reformulated in the model of Croton
et al (2006) to take into account newer observational constraints.

In brief, the model tracks the build-up of the luminous component of
each dark matter halo by prescribing how gas cools and transforms into
stars, as well as how enriched gas is devolved to the interstellar
medium in halos of various masses and at different stages of the
hierarchy. A novel feature of the implementation of Croton et al
(2006) is the inclusion of AGN feedback to curtail star formation in
massive objects and to prevent the formation of overluminous galaxies
at the center of galaxy clusters. The star formation and enrichment
history of galaxies is then processed with standard spectrophotometric
models to provide estimates of galaxy luminosities and colors. The
main data we use here is the luminosity in the 5 Sloan bands,
{\it{ugriz}}, and we concentrate our analysis in the $g$ and $r$
bands. Unless otherwise specified, luminosities and magnitudes will
refer to the $r$ band in what follows.

The semianalytic approach follows explicitly the evolution of galaxies
within dark matter halos, even when they are accreted into (and become
satellites of) larger structures. After accretion, satellite galaxies
are followed consistently until their parent subhalo is destroyed by
the tidal field of the more massive system, at which point the
satellite orbit is subsequently identified as that of the most bound
particle of the parent subhalo before disruption. The satellite is
subsequently assumed to survive for some residual time consistent with
dynamical friction estimates. As we shall see below, this careful
treatment of the N-body simulation is crucial to characterize the
dynamics of satellites of isolated galaxies in the {\it MS}. Some
aspects of the true evolution are nevertheless neglected; in
particular, the effects of dynamical friction on the orbit of a
satellite after its dark halo is disrupted but before it merges with
the central galaxy of its host halo, as well as a necessarily rough
estimation of its merging time. This neglect will affect some aspects
of the results presented below, notably the overall abundance and
the radial distribution of satellites.

\subsection{The Primary/Satellite Galaxy Catalogue}
\label{ssec:gxcat}

\subsubsection{Primary galaxies}
\label{sssec:primgx}

We follow standard practice and identify a sample of primary galaxies
from the {\it MS} galaxy catalogue through (i) an {\it isolation}
criterion, imposed to ensure that a single object dominates the local dynamics
traced by the satellites, and (ii) a brightness cutoff, imposed to
ensure that most primaries have a fair chance of having detectable
satellites. Hereafter, we shall refer to systems brighter than
$M_r=-20.5$ and surrounded, within 1 $h^{-1}$ Mpc, {\it only} by
companions at least 2 magnitudes fainter, as {\it primary
galaxies}. (Note that neither the Milky Way nor M31 satisfy this
strict isolation criterion.)

\subsubsection{Satellites}
\label{sssec:satgx}

We identify as satellites any galaxy brighter than $M_r=-17$ (the
limiting magnitude of the catalogue) that lies within the virial
radius
\footnote{We define the {\it virial} radius, $r_{\rm 200}$, of a
system as the radius of a sphere of mean density $200$ times the
critical density for closure, 
$\rho_{\rm crit}=3H_0^2/8\pi G\sim 277.5 \, h^{2}\, M_\odot/\, \rm kpc^3$.
This implicitly defines the virial mass of a halo, $M_{\rm 200}$,
as that enclosed within $r_{200}$, and the virial velocity, $V_{\rm 200}$, 
as the circular velocity measured at $r_{\rm 200}$.
Quantities characterizing a system will be referred to as ``virial''
and measured within $r_{\rm 200}$, unless otherwise specified.}
of a primary. Primaries with {\it no} satellites brighter than
$M_r=-17$ within their virial radius will be referred to as {\it
singles}. Primaries with at least one satellite will be referred to as
{\it hosts}.

\subsubsection{Further nomenclature}
\label{sssec:nomenc}

One complication arises from the fact that, despite our isolation
criterion, some primaries might themselves be satellites of larger
systems. This tends to happen in the rarefied outskirts of massive
clusters. In this case, the virial radius would refer to the cluster,
rather than to the parent halo of the primary, and the velocities of
nearby satellites would be contaminated by high-speed cluster
members. To take this into account, we shall distinguish, if
pertinent, two classes of primaries: {\it central} and {\it
non-central}. However, we emphasize that the fraction of non-central
primaries is rather small, and that excluding them from the
statistical analysis has no major influence on our main conclusions.

A second note refers to satellites, some of which, as described in
\S~\ref{ssec:sam}, are still surrounded at the present time by their
parent halos (hereafter {\it WSUB} satellites, for short), while
others are identified with the most bound dark matter particle of the
parent subhalo immediately prior to disruption ({\it NOSUB}
satellites). Clearly, results concerning {\it NOSUB} satellites are
likely to be more model-dependent, since the identification of the
satellite as a distinct object depends in this case heavily on the
assumptions made in the model about dynamical friction timescales. It
is important to keep this distinction in mind as we try and interpret
the results.

\subsubsection{Statistics}
\label{sssec:stats}

It is interesting to assess how our primary selection criteria select
galaxies from the general population. We find that the fraction of
galaxies that are designated as primaries depends only weakly on
luminosity: for example, $23\%$ of galaxies brighter than $M_r=-23$
are {\it primaries}, the vast majority of which ($90\%$) are {\it
hosts}; $94.6\%$ of them are of the {\it central} type. Fainter galaxies
($-21<M_r<-20.5$) have a $19\%$ chance of being classified as {\it
primaries}, but the majority of them ($85\%$) are actually {\it
singles} and therefore will not contribute to satellite studies. Of
these fainter primaries, $99.5\%$ are classified as {\it
central}. Overall, only $1\%$ of primaries are {\it non-central}, and
therefore our results are unlikely to be affected by the presence of
these unusual systems.

The final sample contains $258,321$ {\it primaries}, $79,000$ of which
are {\it hosts} to at least one satellite. We find more than $178,000$
satellites orbiting within the virial radius of the primaries, a
number that rises to $508,000$ if all satellites within 1 $h^{-1}$ Mpc
of the primaries are considered. We list the fraction of primaries as
a function of $r$-band luminosity and halo mass, along with other
useful data, in Tables~\ref{tab:mfrac} and ~\ref{tab:lfrac}.

\begin{figure}
\begin{center}
\includegraphics[width=84mm]{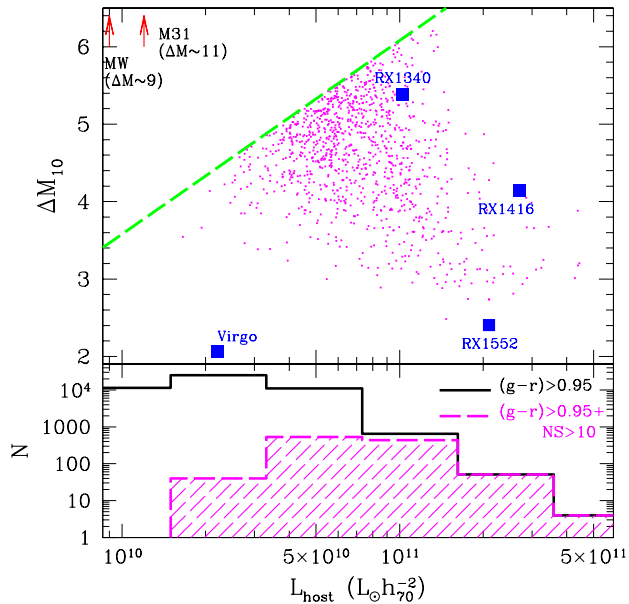}
\caption{ {\it Top panel:} Magnitude gap between the primary and the
10th brightest satellite within $r_{200}$ as a function of
$L_{\rm host}$ (expressed in units of $h_{70}^{-2}\, L_\odot$).  The
dashed line indicates the $M_r=-17$ faint magnitude cutoff of our
catalogue. For reference, we show also the approximate location of a
few well-known systems, such as the Milky Way and M31 (both of which
fall outside the plot), the Virgo cluster \citep{trentham02}, and three ``fossil''
groups: RX J1340.6+4018 \citep{jones00}, RX J1552.2+2013
\citep{mendez_oliveira06} and RX J1416.4+2315 \citep{cypriano06}.
Magnitudes have been converted to the r-band using \citep{fukugita95}
when necessary.
{\it Bottom panel:} Luminosity
distribution of all red ($(g-r)>0.95$) primaries (black solid
histogram).The shaded magenta histogram indicate the subsample shown
in the upper panel. These are the red {\it host} primaries having at
least $10$ satellites within $r_{200}$.}
\label{figs:lgap}
\end{center}
\end{figure}

\begin{center}
\begin{figure}
\includegraphics[width=84mm]{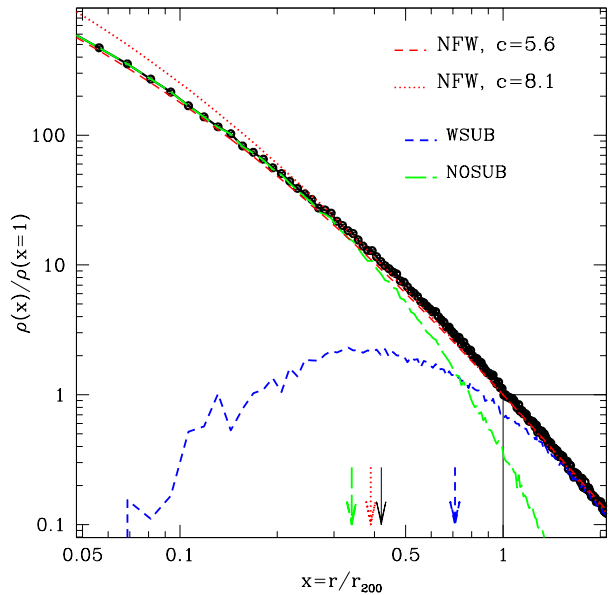}
\caption{Number density profile of satellites within $\sim 2$ virial
radii, computed after scaling the position of all satellites to the
virial radius of the primary halo, and normalized at $r=r_{200}$. The
contribution of {\it WSUB} and {\it NOSUB} satellites to the density
profile is shown as well. Note that the majority of satellites near
the center have had their parent halo stripped. The shape of the
density profile is well approximated by an NFW profile with
$c_{200}\sim 5.6$. This is slightly less concentrated than the
``average'' halo around these primaries, for which we estimate
$\langle c_{200}\rangle \sim 8.1$. Both of these curves are shown in
the figure, using the same normalization as for the satellites. Arrows
indicate the radius containing half of the objects in each profile.}
\label{figs:dprof}
\end{figure}
\end{center}

\begin{center}
\begin{figure}
\includegraphics[width=84mm]{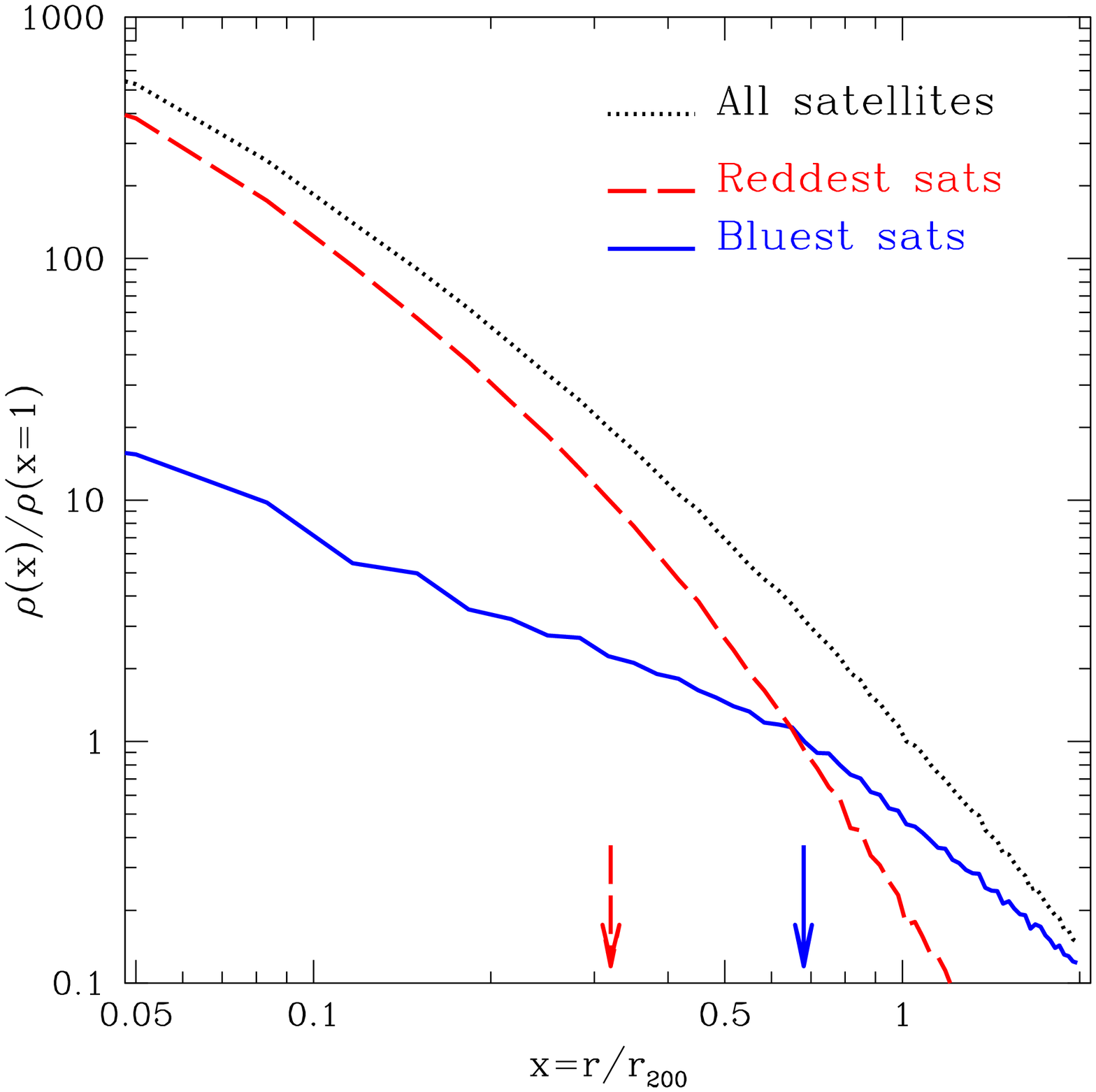}
\caption{The color dependence of the satellite number density
profile. The thick dotted line shows the profile corresponding to {\it
all} satellites, as in Figure~\ref{figs:dprof}. The dashed (red) and
solid (blue) curves show the contribution to the total profile of the
reddest ($(g-r)>1.01$) and bluest ($(g-r)<0.95$) one-third of the
satellites. Note that the reddest satellites are significantly more
concentrated than the bluest ones. Arrows indicate the radius
enclosing half of the satellites in each subsample.}
\label{figs:dprofcol}
\end{figure}
\end{center}

\section {Results and Discussion}
\label{sec:results}

\subsection {Mass-Luminosity-Color relation for primaries}
\label{ssec:lmc}

We begin by exploring the virial mass-luminosity relation for
primaries in our catalogue, one of the main topics addressed by
studies of satellite dynamics. This is shown in the top left panel of
Figure~\ref{figs:lmc}, and illustrates the expected trend for brighter
primaries to inhabit more massive halos. The most striking aspect of
this figure, however, is the large scatter: at given luminosity, halos
span over a decade in mass. Conversely, halos of given mass may host
primaries spanning a factor of $\sim 5$ in luminosity.  The rms
deviation at given $L$ is substantially smaller (see table \ref{tab:lfrac})
but the broad tails
in the mass-luminosity distribution are important, and we shall see
below.  The large scatter in the mass-luminosity relation reflects the
variety of mass assembly and star formation histories of isolated
galaxies, and must be borne in mind when ``stacking'' galaxies
according to luminosity in order to study their average halo
properties.

Different colors in Figure~\ref{figs:lmc} code the number density of
primaries at each location in the $M$-$L$ plane, varying by a factor
of $\sim 10$ for each step in color, and decreasing from red to
blue. Most of the galaxies are in a relatively tight ``main sequence''
(outlined by the red dots in Figure~\ref{figs:lmc} and delineated
roughly by the solid line), but are surrounded by a broad cloud
covering a large fraction of the panel. Closer inspection reveals the
presence of two further relatively well-defined ``sequences'': one
consisting of galaxies that are systematically brighter than the
``main sequence'', indicated roughly by the dotted line, and another
one consisting of ``underluminous'' galaxies inhabiting fairly massive
halos (see dashed line in Figure~\ref{figs:lmc}).

We note that the ``main sequence'' follows roughly a constant virial
mass-to-light ratio. This would be the natural expectation for halos
where a similar fraction of their baryonic content has been
transformed into stars of roughly similar mass-to-light ratio. 

The ``bright sequence'' (dotted line) is caused by the transient
brightening of the luminous component resulting from a starburst
triggered by a major merger.

The ``underluminous sequence'' (dashed line), on the other hand, is
linked to the stunted growth of the stellar component of central
galaxies in massive halos that results from the ``radio-mode'' AGN
feedback (see for details Croton et al 2006). In such systems, halos
can increase their mass substantially while their central galaxies
grow only through mergers. This is indeed the way in which the
semianalytic model is able to reconcile the hierarchical growth of
structure with the observed down-sizing of star formation in galaxies
and the presence of ``red and dead'' galaxies at early epochs (Croton
et al 2006, De Lucia et al 2006, Bower et al 2006). \nocite{bower06,
delucia06}

The above interpretation suggests that the three ``sequences'' may be
best appreciated by applying color cuts to the sample, since starburst
galaxies will be bluer than average, while the opposite will be true
for galaxies that have not formed stars recently. This expectation is
largely borne out, as shown by the other three panels in
Figure~\ref{figs:lmc}: galaxies bluer than $g-r=0.65$ trace
predominantly the ``bright'' sequence, whereas those redder than
$g-r=0.95$ largely trace the ``underluminous sequence''. The ``main
sequence'' is nicely traced by galaxies of intermediate, less extreme,
colors. We hasten to add that the color discrimination is not perfect,
and that residual evidence for the ``main sequence'' is clearly
apparent in both the red and blue panels of Figure~\ref{figs:lmc}.

The main conclusion from this discussion is that, at fixed luminosity,
the virial mass depends strongly on color, and that a strict color
selection criterion applied to primaries helps to tighten
substantially the scaling between virial mass and luminosity 
\citep[see e.g.,][]{klypin_prada07}. As we
discuss below, the large scatter that would result from blindly
stacking primaries in luminosity bins would most likely obscure many
of the underlying trends.

The prevalence of each of these sequences is a sensitive function of
halo mass and galaxy luminosity. This is illustrated in
Figure~\ref{figs:lmcdist}, where each panel shows, for several
luminosity bins, the distribution of virial masses of galaxies in each
of the three sequences. The sample is further split into {\it host}
and {\it single} primaries, since, by definition, only the halos of
{\it host} primaries are amenable to satellite dynamical studies.

Blue primaries are clearly in the minority amongst host galaxies,
except at the brightest magnitudes, where they make a substantial
($\sim 29\%$) fraction of the brightest ($L>10^{11}\, L_\odot$)
primaries. Red galaxies, on the other hand, are reasonably well
represented at all magnitudes, making up about $\sim 30$-$40\%$ of all
primaries fainter than $10^{11} \, L_\odot$. The ``main sequence'' of
primaries dominates at all but the brightest luminosities:
intriguingly, only blue and red primaries make up the population of
$L>10^{11} \, L_\odot$ hosts. These are clearly very unusual objects,
either the result of a recent starburst in a relatively low mass halo,
or the result of growth by mergers but without star formation at the
center of a very massive halo.

{\it Single} primaries show similar mass-color-luminosity trends as
{\it hosts}, although galaxies with extremely red colors are less well
represented, and the ``main sequence'' is more prevalent than amongst
{\it hosts}. Interestingly, only blue {\it single} primaries in low
mass halos populate the brightest luminosity bin ($L>10^{11} \,
L_\odot$): this is clearly the result of a major 
merger fueling a starburst of extreme but short-lived brightness.

At given luminosity, {\it single} primaries tend to have lower virial
masses than {\it hosts}. This is expected, since more massive halos
tend to have more of everything, including satellites, implying a
small bias toward more massive halos in a sample selected to have at
least one satellite. The effect is noticeable, as witnessed by the
shift in the position of the green histogram in corresponding panels
of the {\it single} and {\it host} distributions shown in
Figure~\ref{figs:lmcdist}. As expected from the argument below, it is
largest in the lowest luminosity bins (i.e., $L<10^{10.4} \,
L_\odot$), where it amounts to a shift by a factor of $\sim 2$ in the
average mass of galaxies in the ``main'' sequence.

\subsection {Application to ``fossil groups''}
\label{ssec:fossil}

As is clear from Figures~\ref{figs:lmc} and ~\ref{figs:lmcdist}, our
primary selection criteria select a number of systems in fairly
massive halos, $10^{13}$-$10^{15} \, M_\odot$, a range usually
associated with galaxy groups and poor clusters rather than isolated
galaxies. These are, indeed, systems of galaxies which, by chance,
have evolved a gap of at least two magnitudes between the brightest
and second brightest galaxy, and therefore are included in our
sample. These systems are not rare; according to
Table~\ref{tab:mfrac}, the brightest galaxy of about $\sim 10\%$ of
all halos in the range $10^{13}$-$10^{15} \, M_\odot$ is a primary
according to our definition.

The presence of these systems in our sample has the potential of
biasing our satellite velocities toward large values, and therefore
one must be careful to take this into account in the analysis. We
shall return to this issue below (\S~\ref{ssec:lsg}); but we note here
that these are the analogs of ``fossil groups'', systems of galaxies
which are unusually X-ray bright (and massive) for their optical
richness. ``Fossil'' groups or clusters are defined, just like our
primaries, as systems where a large ($>2$) magnitude gap exists
between the brightest and second-brightest galaxy.

Although the statistics of these objects are still fairly poorly
known, their unusual properties have attracted attention from
observers and theorists alike, and there has been speculation that the
abundance of these systems might be difficult to reproduce in the 
$\Lambda$CDM cosmogony \citep{donghia04,donghia05,milosavljevic06,
sommerlarsen06}. Subsequent work has shown, however, that the predicted
abundance of "fossil" systems is in reasonable agreement with the still
rather uncertain observational constraints \citep[see, e.g.][]{donghia07,
vandenbosch07}. Worries remain that the atypical luminosity function of 
known 'fossil' systems may be difficult to reconcile with the substructure
function of $\Lambda$CDM halos \citep{donghia07}, but we emphasize that 
{\it only three} "fossil" groups have published luminosity functions. This,
however, should change soon, as available X-ray data are systematically 
surveyed for the presence of ``fossil'' groups, and as imaging and 
spectroscopic surveys provide secure luminosity functions for a growing 
number of ``fossil'' groups.

Our intention here is to characterize the abundance and luminosity
function of ``fossil''-like systems in our galaxy catalogue in order
to guide the interpretation of future observational constraints. As
stated above, these objects are not rare, since up to one in ten of
the most massive halos may be regarded as a ``fossil'' system
according to the above definitions (see Table~\ref{tab:mfrac}).  These
systems are not rare amongst bright galaxies either, as shown in
Table~\ref{tab:lfrac}; approximately $22\%$ of galaxies brighter than
$5\times 10^{10} L_\odot$ live in a ``fossil''-like
environment. ``fossil''-like systems are more prevalent amongst
fainter galaxies, which tend to inhabit
lower mass halos, where large magnitude gaps between the two brightest
galaxy members are common (e.g., the Milky Way is the ``ultimate
fossil group'').

This gives a total number density of $8.1 \times 10^{-6}\, h^{3}$
Mpc$^{-3}$ for ``fossil'' systems in halos exceeding $10^{13} \,
h^{-1} \, M_\odot$, or $8.3 \times 10^{-5}\, h^{3}$ Mpc$^{-3}$ for
fossil groups whose brightest galaxy exceeds $5\times 10^{10}
h^{-2}\, L_\odot$. This actually exceeds the (still rather
uncertain) observational estimates of $\sim 5 \times 10^{-7}-2 \times
10^{-6} \, h^{3}$ Mpc$^{-3}$ \citep{jones03, vikhlinin99,romer00} for
groups with X-ray luminosities $\geq 10^{43} h_{50}^{-2}$ erg ${\rm
s}^{-1}$. We note, however, that only about $10$ fossil groups are
actually known, and that this may very well be an underestimate. Our
preliminary conclusion is that the semianalytic approach has no
difficulty accounting for the abundance of groups with such a strong
distinction between brightest and second-brightest galaxy.

A further test concerns the actual luminosity distribution of galaxies
in the groups. In a recent paper, D'Onghia et al (2007) argue, on the
basis of N-body simulations, that ``fossil'' groups present a
challenge to the $\Lambda$CDM scenario, since few cold dark matter
halos are as deficient in substructure as ``fossil'' groups. This
suggestion is based on the analysis of simulations that follow solely
the dark matter component, so it is important to check the predictions
of models that actually follow the formation of the galaxies.

We adopt a simple measure of the shape of the luminosity distribution
of galaxies in a ``fossil'' group, namely the magnitude difference
between the brightest and tenth brightest galaxy member, $\Delta
M_{10}$. This is shown as a function of primary luminosity in the top
panel of Figure~\ref{figs:lgap}. A typical cluster with a well
populated Schechter-like luminosity distribution, such as Virgo, has
$\Delta M_{10} \sim 2$ \citep{trentham02}. Individual galaxies, such as the
Milky Way or M31, have a much larger percentage of the total light
concentrated in a single object, and as a result their $\Delta M_{10}$
differ markedly from Virgo; we find $\Delta M_{10}=9$ and $11$ for the
Milky Way and M31, respectively \citep{mateo98}.
Fossil groups have intermediate values of $\Delta M_{10}$; for 
the three groups with published luminosity functions, the values 
span the range from $\sim 3$ to $\sim 5$ \citep{jones00,
mendez_oliveira06,cypriano06}.

These values are not unusual in our catalogue of isolated primaries,
where $\Delta M_{10}$ is a strong function of the primary luminosity,
approaching $\sim 2$ (the minimum possible value given our isolation
criterion) for the brightest primaries, but increasing rapidly with
decreasing luminosity. As shown in the top panel of
Figure~\ref{figs:lgap}, there is also a large scatter in $\Delta
M_{10}$; at $L\sim 10^{11} L_\odot$ we find values that go from $\sim
3$ to $\sim 6$, spanning the range observed in fossil groups. It would
clearly be difficult to argue on the basis of this evidence that there
is a substantial discrepancy between $\Lambda$CDM predictions and the
observations of ``fossil'' groups. Further data are needed to clarify
this issue further.

\begin{center}
\begin{figure}
\includegraphics[width=84mm]{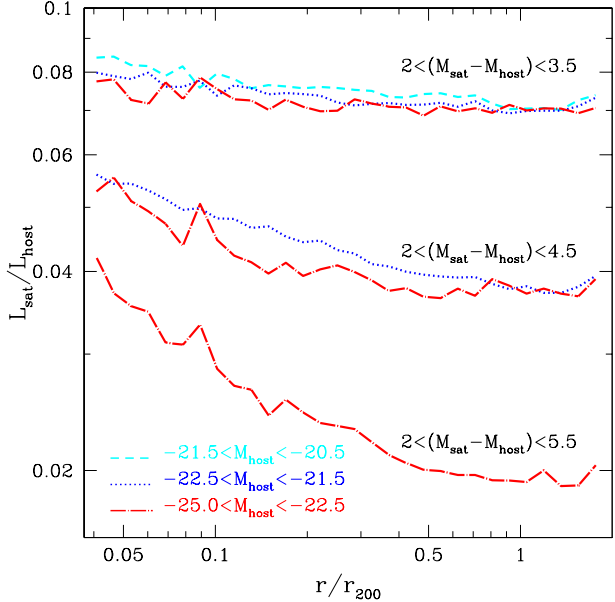}
\caption{The average satellite luminosity (expressed in units of
$L_{\rm host}$) is shown as a function of radius for various primary
luminosity bins. Because of our faint magnitude cutoff ($M_r=-17$),
we consider a different magnitude range for satellites, depending on
the brightness of the host. The three dot-dashed lines correspond to
primaries brighter than $10^{10.8}\, L_\odot$ ($M_r\lsim-22.5$).
Note that when considering the full magnitude range resolved in the {\it MS},
($2<M_{\rm sat}-M_{\rm host}<5.5$) a strong luminosity segregation is
clearly present: the average brightness of satellites decreases by a
factor of $\sim 2$ from the center out to the virial radius. As
expected, the magnitude of the effect decreases as the magnitude range
narrows when considering fainter hosts. Once this effect is taken into
account, the satellite luminosity segregation seems to be roughly
independent of host brightness.}
\label{figs:dprofl}
\end{figure}
\end{center}

\subsection {Satellite density profile}
\label{ssec:satprof}

The solid circles in Figure~\ref{figs:dprof} show the number density
profile of satellites, computed after rescaling the position of each
satellite to the virial radius of the host and stacking the full
sample. The large number of satellites in our catalogue makes the
statistical error in the profile negligible; bootstrap error bars are
smaller than the size of each symbol.

The shape of the satellite density profile is well described by the
NFW (Navarro, Frenk \& White 1996, 1997) formula, with
$c_{200}=5.6$. This result is rather insensitive to halo mass or
luminosity, for example, splitting the sample of primaries in two by
halo mass results in $c=5.6\pm0.3$ and $6.2\pm0.5$ for the high and 
low mass sample, respectively. A similar split in luminosity yields 
$c=6.2 \pm 0.6$ and $c=5.5 \pm 0.4$, respectively. 
\nocite{nfw96,nfw97}

Satellites are slightly less concentrated than the host dark matter
halos.  The average host dark halo concentration may be found by averaging
the concentrations of host halos, estimated from the
mass-concentration relation of Fausti-Neto et al (in preparation):
\begin{equation}
\log(c_{200})= 2.1-0.1 \, \log(M_{200}/(M_\odot/h)^{-1}),
\end{equation}
and taking into account the lognormal dispersion
$\sigma(\log(c_{200}))=0.10$. We find an average concentration of
$\langle c_{200} \rangle =8.1$.

These two NFW profiles are plotted in Figure~\ref{figs:dprof}, showing
 that the difference is not large. For example, the half mass radius
 of the average dark halo is $0.39 \, r_{200}$, which is very similar
 to the radius that contains half of the stacked satellites, $0.42 \,
 r_{200}$. We conclude that satellites are a relatively unbiased
 tracer of the dark mass distribution within the virial radius of a
 halo, at least for this model of galaxy formation and evolution.

In a recent paper, \citet{chen06} study the radial projected
distribution of satellite galaxies in the SDSS. These authors find
that the projected satellite number density profile is well fitted by
a power law: $\Sigma(R) \propto R^\alpha$, with
$\alpha=-1.7\pm0.1$. In order to approximately mimic the selection
criteria applied by Chen et al., we have projected all galaxies in our
catalogue within 1 $h^{-1}$Mpc around each isolated primary,
considering as "satellites" those within projected distances $\Delta
R<500\; h^{-1}$ kpc and line-of-sight velocities $\Delta V_{los}<500$
km/s (notice that from our definition of isolated galaxies, all
"satellites" are at least two magnitude fainter than the primary). We
find $\alpha=-1.55 \pm 0.08$ in the distance range $26<R<500 \,
h^{-1}$kpc, which is consistent with the results of Chen et al within
the quoted errors.

\subsubsection{Dependence on satellite color}
\label{sssec:dprofcol}

Although the satellite population as a whole traces the dark mass
reasonably well, there is a strong dependence on satellite color. This
is shown in Figure~\ref{figs:dprofcol}, where the profiles of the
reddest ($(g-r)>1.01$, red dashed line) and bluest ($(g-r)<0.95$, solid
blue line) one-third of satellites is compared with the overall
profile presented in Figure~\ref{figs:dprof}. Red satellites are
clearly much more centrally concentrated than blue ones: half of the
red sample is contained within $0.32 \, r_{200}$, a radius that climbs
to $0.71 \,r_{200}$ for the bluest one-third of satellites.

This difference in concentration may be traced to the assumption in
the semianalytic treatment that a satellite loses its reservoir of hot
gas (the future fuel for star formation) once it is accreted into a
larger structure. Thus star formation in satellites declines quickly
after accretion: the earlier a satellite was accreted the older (and
redder) its stellar population will be. Early-accreting satellites
have smaller turnaround radii and are likely to orbit closer to the
center than late accreting ones, resulting in the trend shown in
Figure~\ref{figs:dprofcol}. 

A trend of similar origin is shown also by the bottom two curves in
Figure~\ref{figs:dprof}, which show the contribution to the satellite
density profile from satellites that have preserved ({\it WSUB}) or
lost ({\it NOSUB}) their parent dark halos. Early accreting satellites
are more likely to have been more affected by tides and to have lost
their parent halos, leading to the spatial segregation between {\it
WSUB} and {\it NOSUB} satellites seen here. This figure also shows
clearly the importance of tracking satellites in dark matter-only
simulations even after their parent halos have been disrupted in the
tidal field of the primary: {\it NOSUB} satellites are crucial to the
satellite profile in the inner regions of the primary.

\begin{center}
\begin{figure}
\includegraphics[width=84mm]{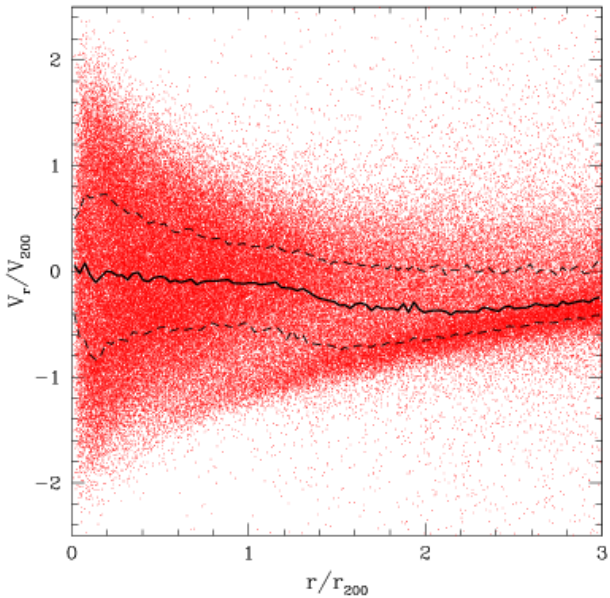}
\caption{The radial velocity of all satellites in our sample as a
function of radius, both scaled to the virial values of the host
halo. The solid curve shows the median of the distribution, while
dashed lines outline the 25\% and 75\% percentiles of the distribution
as a function of radius. The velocity dispersion decreases out to the
virial radius beyond which it remains approximately constant. A
``first infall'' sequence of negative radial velocities is clearly
defined outside $\sim 0.5 \, r_{200}$. Satellites with very high
radial velocities (sometimes exceeding $\sim 3\, V_{200}$) are present,
especially at large radii. This is a result of some primaries lying in
the periphery of much larger structures, such as galaxy clusters; the
large velocities are those of cluster members typically unrelated to
the primary.}
\label{figs:rvr}
\end{figure}
\end{center}

\subsubsection{Dependence on satellite brightness}
\label{sssec:dprofl}

Figure~\ref{figs:dprofl} explores the luminosity segregation of
satellites and its dependence on primary luminosity. The faint
magnitude cutoff of our catalogue implies that, in order to compare
meaningfully primaries of different brightness, satellites must be
selected within a definite magnitude range. Satellites can only have
absolute magnitudes between the cutoff at $M_r=-17$ and $M_{\rm
host}+2$, so that fainter primaries have, by construction, satellites
that span a narrower magnitude range.

The three (red) dot-dashed curves in Figure~\ref{figs:dprofl} indicate
the radial dependence of the average satellite luminosity (in units of
the host's) for the brightest hosts ($L_r>10^{10.8} \, L_\odot$;
$M_r<-22.5$). When all
satellites are considered (i.e., in the magnitude range $2<M_{\rm
sat}-M_{\rm host}<5.5$; bottom curve in Figure~\ref{figs:dprofl}) a
significant radial trend is seen: the average satellite luminosity
drops by a factor of $\sim 2$ from the center out to the virial
radius. The trend is, as expected, more difficult to detect when a
narrower range in satellite brightnesses is imposed (upper two
curves). Such a narrower range is needed when considering fainter
primaries, for which results are shown by the dashed and dotted
curves in Figure~\ref{figs:dprofl}. Once this is taken into account
the luminosity segregation we find seems to be independent of the
brightness of the primary.

The luminosity segregation shown in Figure~\ref{figs:dprofl} is most
likely due to the effects of dynamical friction, which operate faster
on more massive/more luminous satellites, bringing them closer to the
center on a shorter timescale. It should be possible to contrast this
result against observation.  However, a note of caution about this
interpretation is in order, recalling the results of
Figure~\ref{figs:dprof}. Since most satellites, especially those near
the center, have lost their parent halo, their present-day location is
being traced by a single dark matter particle, and their survival
depends directly on the semianalytic model assumptions about dynamical
friction. As such, the luminosity segregation shown in
Figure~\ref{figs:dprofl} is likely to be model-dependent.

\subsection {Satellite Velocities}
\label{ssec:satvel}

\subsubsection{Radial velocities}
\label{sssec:radv}

Figure~\ref{figs:rvr} shows the radial velocities of all satellites in
our sample as a function of their distance to the primary, after
rescaling to the virial quantities of the host halo. The solid line
traces the median satellite velocity in radial bins; the dashed lines
the 25\% and 75\% of the distribution, respectively. Note that the
median velocity within $r_{200}$ is constant and consistent with zero,
as expected from a relaxed population in equilibrium.

Outside $r_{200}$, negative velocities are more prevalent, as
satellites on their first approach to the primary start to
dominate. These ``first-infall'' satellites delineate the negative
velocity boundary at all radii, forming a sequence that becomes fairly
obvious outside $\sim 0.5\, r_{200}$ in Figure~\ref{figs:rvr}. The
velocity along this sequence decreases outward, and approaches zero at
$r \sim 3\, r_{200}$, the approximate location of the
turnaround radius according to the simple secondary infall model
\citep[see][]{bertschinger85,white93,navarroandwhite93}.

The ``first-infall'' sequence is an interesting feature of the radial
velocity distribution, one whose detection may be used to yield a
direct estimate of the mass of the host halo. This is the rationale of
a number of studies that attempt to pin down the location of the
turnaround radius in the outskirts of galaxy groups and clusters by
looking at ``caustics'' in the velocity distribution
\citep{diaferio97,diaferio99,geller99,biviano03,rines03,diaferio05}.
Although some progress has been made
on this issue, the observational evidence remains elusive and its
interpretation controversial \citep{reisenegger00,drinkwater01,
mahdavi05,mohayaee06a,gavazzi06}.

Figure~\ref{figs:rvr} offers a possible explanation for these findings:
satellites on first approach make up a relatively small fraction of
systems populating the outskirts of the halo. We investigate this
quantitatively in Figure~\ref{figs:rvdist}, where we show the
satellite radial velocity distribution as a function of radius,
$x=r/r_{200}$, normalized to the virial radius. 

Satellites within the
virial radius have an approximately Gaussian velocity distribution
with dispersion that declines with radius. Interestingly, the radial
velocity dispersion in the inner regions is comparable to the virial
velocity of the halo, $\sigma_r\sim 0.96 \, V_{200}$. This result has
been used by Sales et al (2007a) to argue that the relatively low
velocity dispersion of Galactic satellites implies a fairly low mass
for the Milky Way halo ($V_{200}^{MW}\sim 110$ km/s, see that paper for
further details). The velocity dispersion declines outward: the best
fitting Gaussian for satellites with $0.5<x<1$ is $\sigma_r\sim 0.7\,
V_{200}$.

Beyond the virial radius, the radial velocity distribution becomes
clearly asymmetric, with an excess of satellites with negative radial
velocities. As mentioned above, this is a result of the increasing
importance of satellites on their first approach to the host halo. We
have chosen to quantify this with a double Gaussian fit: one of zero
mean velocity and dispersion fit to the distribution of {\it positive}
radial velocities, and a second one whose dispersion and mean are
chosen so that the sum of the two Gaussians match best the whole
distribution. The two Gaussian fits are shown with dashed lines in
Figure~\ref{figs:rvdist}, and we shall hereafter refer to them, for
short, as the ``zero mean velocity'' component ({\it zmv}) and the
``infall'' component ({\it inf}). Values quoted in each panel of
figure \ref{figs:rvdist} indicate the best-fit Gaussian parameters to
the {\it total} distribution\footnote{We will use $\bar V_i$ and
$\sigma_i$ to indicate the mean and dispersion of the $i$-component
velocity distribution, where $i=r,\theta,\phi$. The skewness $\xi$ is
defined as $\frac{\Sigma (V_i-\bar V_i)^3}{(N-1)\sigma_i^3}$ and the
kurtosis $\kappa=\frac{\Sigma (V_i-\bar V_i)^4}{(N-1)\sigma_i^4} -3$}.
In contrast with the results of \cite{prada06} and \cite{power06}, the
infall patern is insensitive to halo mass; we find a non-negligible
population of first-infall satellites around low and high-mass halos
in our catalogue.

As expected, the mean infall velocity decreases outward, from ${\bar
V}_r \sim -0.85 \, V_{200}$ at $1<x<1.5$ to ${\bar V_r}\sim
-0.4\,V_{200}$ at $2.5<x<3$, and consistent with a turnaround radius
located just outside $\sim 3\, r_{200}$. Interestingly, the velocity
dispersion of infalling satellites amounts to about $15\%$ and $25\%$
of the virial velocity of the host halo, about a factor of two to
three colder than the rest of the population. The prevalence of the
first-infall population increases outward: it makes up $15\%$, $25\%$,
$36\%$, and $40\%$ of all satellites in each of the four $x>1$ bins
shown in Figure~\ref{figs:rvdist}, respectively. These results may be
used to improve algorithms intended to detect infalling galaxies in
the regions surrounding groups and clusters.

\begin{center}
\begin{figure}
\includegraphics[width=84mm]{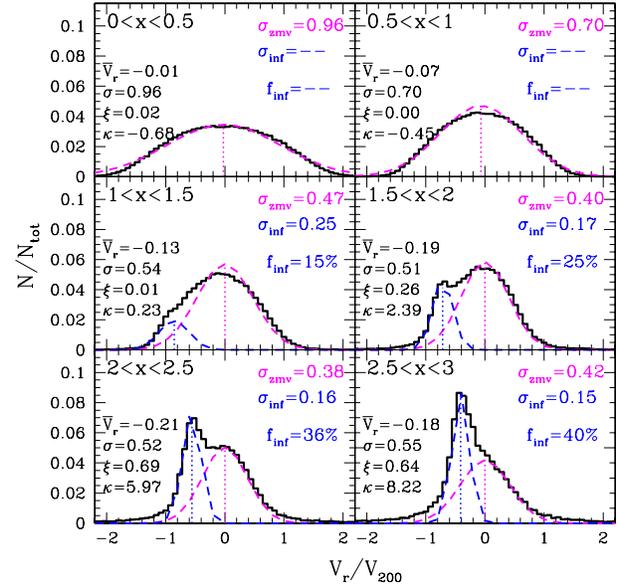}
\caption{Distribution of satellite radial velocities, in bins of
different distance to the primary. We consider only {\it central}
hosts here in order to minimize the contribution of velocity
interlopers. The radial range considered in each panel is labeled by
the range in $x=r/r_{200}$ used. Dashed lines show Gaussian fits to
the profiles: {\it two} Gaussians are used when the distribution shows
a strong asymmetry between negative and positive radial
velocities. One of the Gaussians (the {\it zmv} component) is assumed
to have zero mean velocity and to match the distribution of positive
radial velocities; the parameters of the second one (the {\it inf}
component) are then fit so as to match the whole distribution. The fit
parameters (mean $V_r$, dispersion $\sigma_r$, skewness $\xi$ and kurtosis
$\kappa$) are given in Table~\ref{tab:Gaussfits} as well as quoted 
in each panel.}
\label{figs:rvdist}
\end{figure}
\end{center}

\begin{center}
\begin{figure*}
\includegraphics[width=0.475\linewidth,clip]{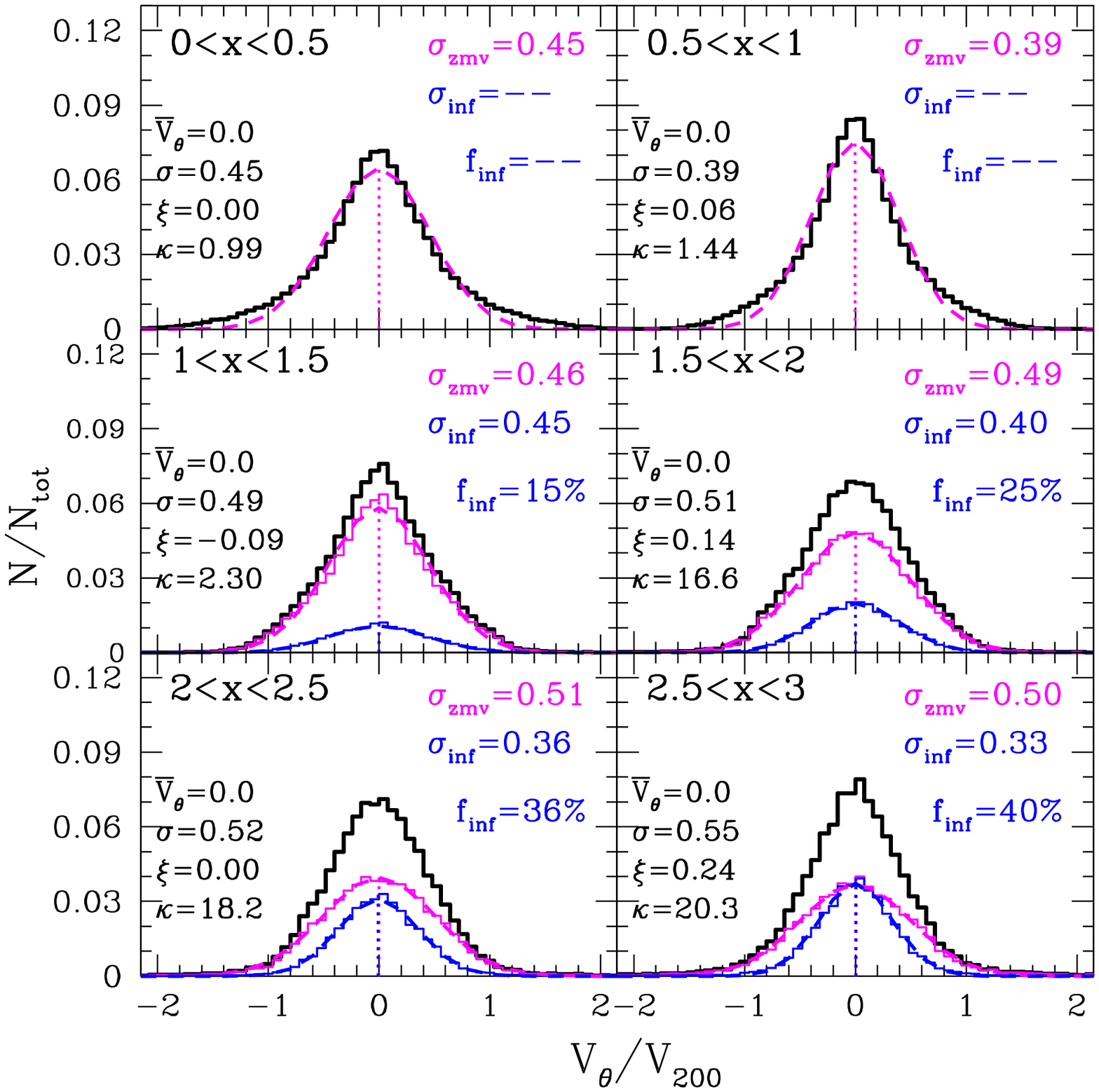}
\hspace{0.4cm}
\includegraphics[width=0.475\linewidth,clip]{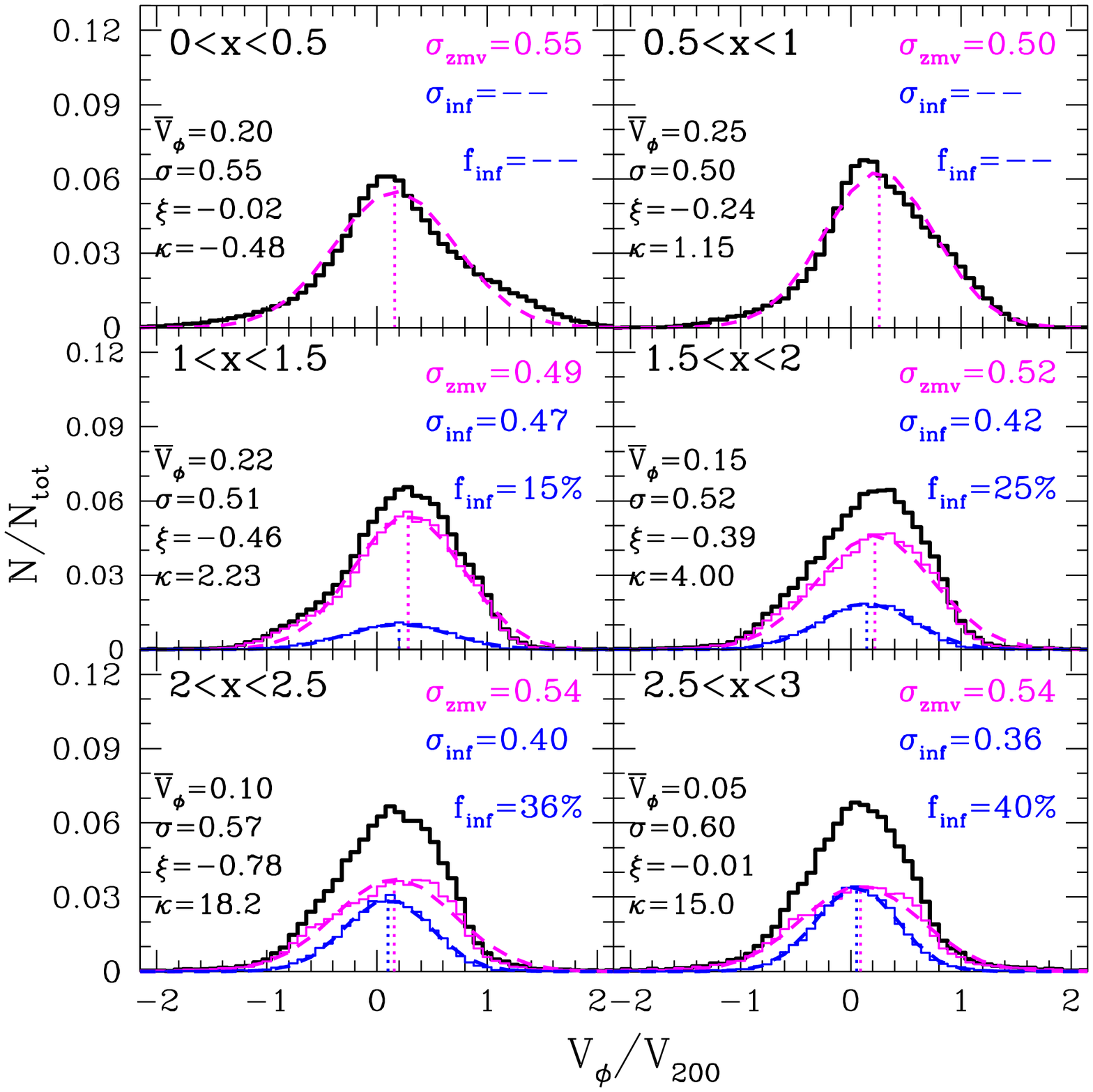}
\caption{Distribution of the tangential spherical components of the
satellite velocities, $V_\theta$ and $V_\phi$, shown in various
distance bins, $x=r/r_{200}$, as labelled in each panel. We have split
the sample in two components, an ``infall'' component ({\it inf},
magenta long-dashed lines) and a ``zero-mean velocity'' component
({\it zmv}, solid black line), using the double Gaussian fits to the
radial velocity distribution of Figure~\ref{figs:rvdist} to assign
probabilistically satellites to each. Dotted curves are Gaussian fits
to each of these distributions. Parameters (mean, dispersion $\sigma$,
skewness $\xi$ and kurtosis $\kappa$, are quoted in each panel 
and in Table~\ref{tab:Gaussfits}.}
\label{figs:vtandist}
\end{figure*}
\end{center}
\subsubsection{Tangential velocities}
\label{sssec:vtan}

In order to complete our characterization of satellite velocities we
show in Figure~\ref{figs:vtandist} the distribution of the tangential
velocity components, binned by distance just as in
Figure~\ref{figs:rvdist}. The spherical components are measured in a
reference frame where the $z$-axis is chosen to coincide with the
angular momentum of the host halo, so that $V_{\phi}$ carries
information about a satellite's sense of rotation relative to the
host.

Within the virial radius the dispersion in $V_{\phi}$ and $V_{\theta}$
is substantially lower than the radial dispersion, and there are also
hints of significant departures from Gaussianity. There is, for
example, an excess of satellites that co-rotate with the host (i.e.,
$V_{\phi}>0$); also, the $V_{\theta}$ distribution is platykurtic;
i.e., it is more centrally peaked and has broader wings than a
Gaussian. Such departures from Gaussianity are more pronounced for
satellites outside the virial radius.

The long- and short-dashed curves in Figure~\ref{figs:vtandist} show
the contribution to the total velocity distribution of the ``infall''
and ``zero-mean velocity'' components discussed in
\S~\ref{sssec:radv}. This is done by assigning satellites
probabilistically to each of the two components, according to the
Gaussian decomposition shown in
Figure~\ref{figs:rvdist}. Interestingly, the $V_{\theta}$
distribution, which is {\it not} well approximated by a Gaussian, is
found to be the result of adding the nearly Gaussian distributions
corresponding to the ``infall'' and the ``zero-mean velocity''
components. Note, as well, that even {\it inside} $r_{200}$ the
$V_{\theta}$ distribution differs slightly from a Gaussian. This is
most likely due to the presence of infalling satellites within the
virial radius. Although they are difficult to pick up in radial
velocity, there are apparently enough of them to modify the
$V_{\theta}$ distribution from Gaussian to platykurtic.

As may be seen from the right-hand panel of
Figure~\ref{figs:vtandist}, the $V_{\phi}$ distribution of the
``infall'' component is approximately Gaussian as well, and shows only
a weak excess of satellites co-rotating with the host. The ``zero-mean
velocity'' component, on the other hand, shows a much more pronounced
asymmetry and a broader dispersion. It is not clear at this point what
causes this difference, but we plan to follow it up in future work. In
all cases, the dispersion in the ``infall'' component is significantly
lower than in the ZMV component. Table~\ref{tab:Gaussfits} lists a
summary of the fit parameters for all these velocity components.

\subsubsection{Velocity anisotropy}
\label{sssec:vanis}

The velocity dispersion of the various spherical components declines
gradually with radius, as shown in Figure~\ref{figs:vanis}. The
biggest decline is seen for the radial velocity dispersion, which
drops by almost a factor of two from the center out to the virial
radius. The dispersion in the other components drops with radius at a
different rate, leading to an anisotropy profile that increases from
the center outwards, reaches a maximum of $\beta_{\rm max}\sim 0.5$ at
$r\sim 0.2\, r_{200}$ and declines to become almost isotropic
$\beta\sim 0$ just outside the virial radius.

The radial dependence of the anisotropy is a reflection of the
increasing importance of the first-infall population at larger
radii. Because it is on its first approach, the {\it inf} population
is ``stretched'' along the radial direction and has therefore a
smaller radial velocity dispersion at any given radius, compared with
its tangential dispersion (compare, e.g., the radial and tangential
dispersions for the {\it inf} population in Figures~\ref{figs:rvdist}
and ~\ref{figs:vtandist}). As the prevalence of this component
increases outwards, it brings down the radial bias characteristic of
the inner regions, leading to a decline of the anisotropy in the
outskirts of the system. In support of this interpretation, we note
that the anisotropy of {\it WSUB} satellites differs strongly from
that of {\it NOSUB} satellites (Figure~\ref{figs:vanis}). The former
still retain their parent halos; are therefore more likely to have
been accreted into the system more recently; and have velocities 
that are more isotropic than the rest.

Figure \ref{figs:vanis} shows that the satellite velocity
dispersion at the virial radius is about $40\%$ its maximum value
in the inner regions. This decline is in agreement with that expected
for NFW halos and can be successfully recovered from observational
samples when contamination from interlopers is properly accounted for.
Although early studies suggested a nearly flat satellite velocity
dispersion profiles favouring isothermal models for host halos
\citep{mckay02,brainerd03}, subsequent analysis suggested that it might
be due to poor removal of background and foreground interlopers. \citet{prada03} studied
the dynamic of satellite
galaxies in the SDSS removing interlopers from their samples fitting
a ''Gaussian + constant'' function to the satellite velocity distributions.
These authors find that the projected velocity dispersion of {\it true}
satellites drops to $\sim 40-60\%$ its maximum value at a projected
distance of $\Delta R\sim 300 h^{-1}$kpc from the primary. The analysis
of satellites in the 2dFGRS catalogue also show a declining $\sigma$
profile after interloper remotion, although the measured drop in the
velocity dispersion with distance is somewhat weaker \citep{brainerd04b}.
The satellite galaxy kinematics in the Millennium Simulation appears broadly consistent
with these observational results.

\subsection{Halo mass profile from satellite dynamics}
\label{ssec:mprof}

Once the spatial distribution and the kinematics of the satellite
population have been characterized, we may use them to constrain the
shape of the host halo mass profile. Assuming spherical symmetry,
equilibrium, and that satellites are massless tracers of the
potential, Jeans' equations link the potential with the velocity
dispersion and density profiles of the satellites. Expressed in terms
of a circular velocity, we have:

\begin{equation}
V_c^2(r)=-\sigma_r^2 \,
(\frac{d\ln{\rho}}{d\ln{r}}+\frac{d\ln{\sigma_r}^2}{d\ln{r}} + 2\beta),
\label{eqs:jeans}
\end{equation}
where the terms in the right hand side may be estimated from the
results in the preceding discussion, and are summarized in the top
panel of Figure~\ref{figs:mprof}.

The implied circular velocity profile for the {\it average} halo
populated by the satellites is shown in the bottom panel of
Figure~\ref{figs:mprof}. This $V_c$ profile has several of the same
characteristics of the NFW profile: it rises to a maximum and then
drops near the virial radius. The maximum circular velocity implied is
$V_{\rm max}\sim 1.25 \, V_{200}$, and occurs at $r_{\rm max}\sim 0.3
\, r_{200}$, which corresponds to a concentration of $c_{200}\sim
7$. We note that this is higher than the concentration ($\sim 5.6$)
derived from fitting an NFW profile to the number density of
satellites (Figure~\ref{figs:dprof}), and is closer to the {\it
average} concentration of $\langle c \rangle \sim 8.1$ found for the
host halos in the {\it MS} (see \S~\ref{ssec:satprof}). In spite of
these differences, it is remarkable that satellites are overall
reasonably good tracers of the dark mass profile. These similarities
have also been reported in N-body/gasdynamical simulations by Sales et
al (2007a), and augur well for studies of the mass profile of galactic
halos based on satellite data.

\begin{center}
\begin{figure}
\includegraphics[width=84mm]{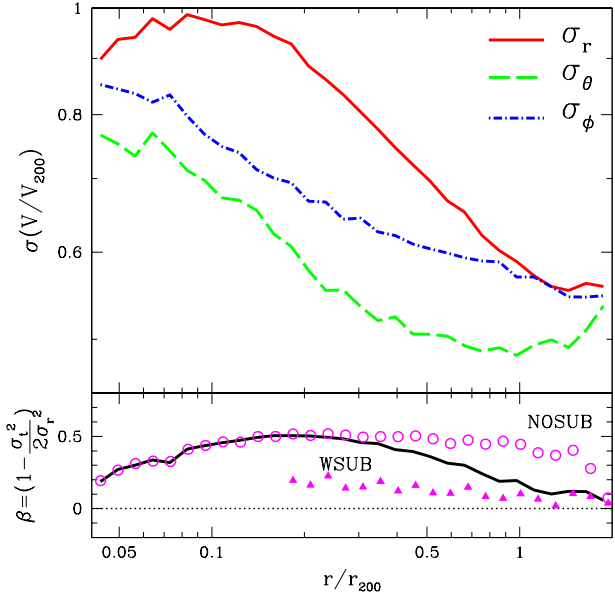}
\caption{{\it Upper panel:} Velocity dispersion profile of central
host satellites in spherical coordinates, where the polar $z$ axis is
chosen to coincide with the direction of the host angular momentum.
{\it Lower panel:} The corresponding anisotropy parameter,
$\beta=1-{\sigma_t}^2/2{\sigma_r}^2$, where
${\sigma_t}^2=\sigma_\theta^2 + \sigma_\phi^2$.  We distinguish in the
bottom panel the contribution of satellites that have preserved their
parent dark matter halo ({\it WSUB}, filled triangles) and those that
have not ({\it NOSUB}, open circles).}
\label{figs:vanis}
\end{figure}
\end{center}

\begin{center}
\begin{figure}
\includegraphics[width=84mm]{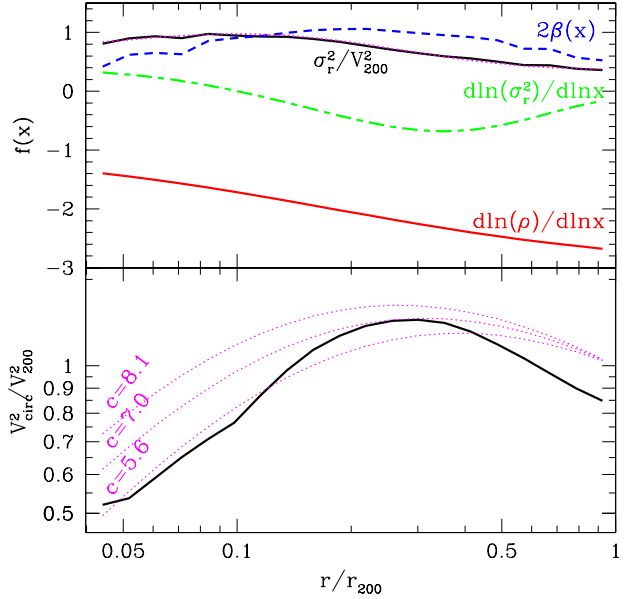}
\caption{{\it Top:} Terms in the right-hand side of Jeans' equation
relating satellite dynamics and the halo mass profile
(eq.~\ref{eqs:jeans}). The dotted line shows a fit to $\sigma_r(r)$ of
the form: $\sigma_r/V_{200}=\sigma_0+ (x/x_0)\, \exp(-(x/x_0)^\alpha)$, with
$x=r/r_{200}$. Best-fitting dimensionless parameters are: $x_0=0.068$,
$\alpha=3/4$ and $\sigma_0=0.6$. 
{\it Bottom:} Solid line shows the average circular
velocity profile of the potential sampled by the satellites, as
derived from applying Jeans' equation. Note that the circular velocity
implied by the satellite population rises to a maximum before dropping
near the virial radius. The maximum circular velocity implied is
consistent with an NFW profile with $c_{200}\approx 7$. This is in
reasonable agreement with the average concentration ($\langle c
\rangle \approx 8.1$) of host halos inhabited by satellites, and is
higher than the concentration derived from the satellite density
profile in Figure~\ref{figs:dprof}.}
\label{figs:mprof}
\end{figure}
\end{center}

\subsection {Primary luminosity vs satellite velocity dispersion}
\label{ssec:lsg}

The velocity dispersion of satellites is one of the prime tools used
to investigate the mass of dark halos surrounding isolated galaxies
and its dependence on luminosity.  In hierarchical formation scenarios
like $\Lambda$CDM, velocities vary with mass according to the virial
definitions discussed in the footnote to \S~\ref{sssec:satgx}, which
imply that mass scales with velocity like $M_{200}\propto V_{200}^3$.

Due to the small number of satellites surrounding each primary, it is
necessary to combine them in some way in order to beat small-number
statistics. One obvious choice is to bin primaries in a narrow
luminosity range, stack their satellites, compute their velocity
dispersion, and see how the dispersion varies as a function of the
luminosity of the primary.

If satellite velocities are an unbiased tracer of the virial velocity,
and luminosity is a reasonable proxy for mass, then one may expect
$L\propto \sigma^3$.  The former assumption appears to be borne out by
the analysis presented above, but the latter one is afflicted by the
large scatter in the mass-luminosity relation discussed in
\S~\ref{ssec:lmc}. How does this affect the scaling between primary
luminosity and satellite velocity dispersion?

We show this in Figure~\ref{figs:lsg}, where hosts are binned
according to their $r$-band luminosity, and the one-dimensional
velocity dispersion is computed for each bin after stacking all of
their satellites. In each panel, the solid circles show the result of
this procedure including all satellites of hosts satisfying the
condition expressed in the label. The short-dashed (blue), long
dashed (green) and dot-dashed (red) curves correspond to
selecting primaries according to the color cuts adopted in
Figure~\ref{figs:lmc}. The dotted lines indicate, for reference, the
$L\propto \sigma^3$ and $L\propto \sigma^2$ scalings.

When considering all primaries (top-left panel), the $L$-$\sigma$
relation is poorly fit by a power law, bending from approximately
$L\propto \sigma^2$ at faint luminosities to a substantially shallower
scaling for $L>5\times 10^{10} L_\odot$.
A similar departure of the $L-\sigma$ relation from a simple power-law scaling has been
reported and discussed previously by \citet{vandenbosch04}.
 This is quite 
different from the naive scaling mentioned above for hierarchical 
models, and is largely due to the large scatter in the halo 
mass-luminosity relation shown in Figure~\ref{figs:lmc}. The presence
of some very massive halos at all luminosities, together with their
growing importance with increasing luminosity bend the relation off the natural $L\propto
\sigma^3$ scaling. At $\sim 5 \times 10^{10} L_\odot$ the prevalence
of such massive objects increases and the relation becomes even
shallower.

\begin{center}
\begin{figure}
\includegraphics[width=84mm]{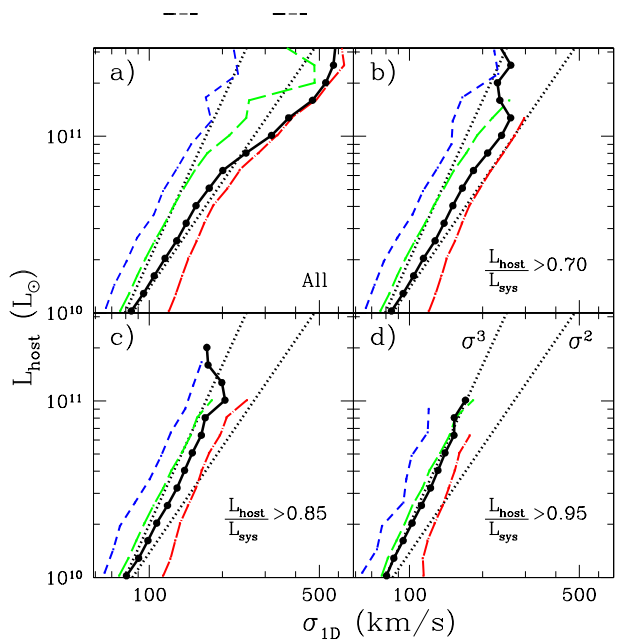}
\caption{Velocity dispersion of satellites around isolated galaxies
binned by $r$-band luminosity. Dispersions are computed after stacking
all satellites within the virial radius of the host of each galaxy in
the bin. Solid connected circles show the results for all satellites
of primaries satisfying the criterion indicated by the label in each
panel. Short-dashed (blue), long-dashed (green) and dot-dashed (red)
curves correspond to primaries selected according to the color cuts
chosen in Figure~\ref{figs:lmc}. Dotted lines indicate $L\propto
\sigma^3$ and $L\propto \sigma^2$ scalings, to ease the comparison
from panel to panel. Panels differ in the sample of primaries
used. The importance of the primary within its own virial radius
distinguishes these samples, as given by $L_{\rm host}/L_{\rm sys}$,
where $L_{\rm sys}$ is the total luminosity of galaxies within the
virial radius. }
\label{figs:lsg}
\end{figure}
\end{center}
\begin{center}
\begin{figure}
\includegraphics[width=84mm]{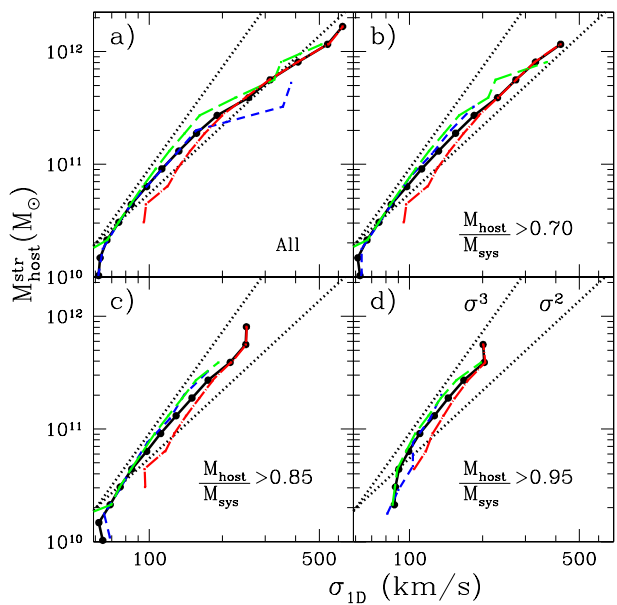}
\caption{Same as figure \ref{figs:lsg}, but considering stellar
masses rather than $r$-band luminosities. Cuts applied on 
$M_{\rm host}/M_{\rm sys}$ refer also to the mass in stars.}
\label{figs:mstsg}
\end{figure}
\end{center}

Restricting primaries by color helps; for example, for blue hosts
($(g-r)<0.65$) we obtain a $L\propto \sigma^3$ relation that holds
over the whole luminosity range.  However, color cuts work less well
for redder primaries: for $0.65<(g-r)<0.95$ hosts we recover the
$L\propto \sigma^3$ relation at faint luminosities, although for
galaxies brighter than $\sim 6 \times 10^{10} L_\odot$ the scaling
becomes again much shallower as a result of the growing importance of
massive halos with underluminous central galaxies. The same aplies to
the primaries with the reddest colors, although the change in the
slope occurs at fainter luminosities ($\sim 4 \times 10^{10}
L_\odot$).  Selecting by color alone is thus not enough to ensure a
sample of primaries with a well-defined power-law scaling between
luminosity and velocity dispersion.

One way of eliminating underluminous primaries within very massive
halos from our sample is to consider the richness of their
surroundings. Massive halos will typically host a large number of
galaxies and in such systems, despite our isolation criteria, which
ensures the dominance of the central galaxy, the other members may
contribute a significant fraction of the total luminosity. This is
shown in the other three panels of Figure~\ref{figs:lsg}, where we
consider only primaries making up more than, respectively, $70\%$,
$85\%$, and $95\%$ of the total combined luminosity of galaxies within
the host halo, $L_{\rm sys}$. The stricter the criteria for selecting
the primaries the nearer the scaling is to the ``natural'' $L\propto
\sigma^3$ relation. Selecting truly isolated galaxies thus requires
more than just imposing a magnitude gap, but also a color cut and a
conscientious survey of the surroundings to weed out fossil groups and
poor clusters from the sample which may unduly bias the $L$-$\sigma$
scaling. Alternatively, one may try and eliminate from the
analysis the brightest galaxies, where the contamination by ``fossil''
systems is worse. This is the approach adopted by \cite{prada03}, who
show that the ``natural'' scaling may also be recovered in that case.

Despite the pruning of the sample, the dependence of the satellite
velocity dispersion on the color of the primary remains. For our
strictest isolation criterion, $L_{\rm host} > 0.95 \, L_{\rm sys}$,
the velocity dispersion of satellites of red galaxies is $\sim 55\%$ higher
than that of blue galaxies of given luminosity. This is reminiscent of
the observations of Brainerd T. (2004a,b) in the 2dFGRS
who found larger velocity dispersion in satellites associated to 
early-type (red) primaries than satellites of late-type (blue)
hosts.\nocite{brainerd04a,brainerd04b}

This effect is due largely to the different stellar mass-to-light
ratios of galaxies of different colors. Indeed, as shown in
Figure~\ref{figs:mstsg}, the shift in velocity dispersion between
primaries of different color basically disappears when {\it stellar
masses} are considered instead of $r$-band luminosities. Using,
whenever possible, stellar mass estimates rather than luminosity in
order to bin galaxies is likely to give more robust results. Combining
this with a strict isolation criterion that evaluates not only the
luminosity gap between brightest and second-brightest galaxy but also
the richness of the surrounding field appears essential in order to
avoid biases and to recover the natural scaling expected from
hierarchical structure formation scenarios.

\subsection {Spatial anisotropies and the Holmberg effect}
\label{ssec:holm}

An issue that has drawn recurring attention over time is whether the
spatial distribution of satellites around primaries has anisotropies
of particular significance. For example, the brightest satellites
around the Milky Way seem to align on a plane perpendicular to the
disk of the Galaxy \citep{lyndenbell82,majewski94,kroupa05,libeskind05},
and a similar result seems to apply to
at least some of the satellites of the Andromeda galaxy
\citep{kochandgrebel06,metz07}. The small number of
satellites involved in these analyses precludes robust conclusions to
be drawn on the basis of the Local Group \citep{zentner05}, but
it is intriguing that both Holmberg (1969) and \citet{zaritsky97b}
find a similar effect in their samples of satellites of spiral
galaxies.

The advent of large datasets, mostly from the SDSS, has allowed the
issue to be revisited, and the latest work suggests that satellites of
isolated spirals tend to distribute themselves preferentially along
the direction of the disk: the {\it opposite} of the effect claimed by
Holmberg and present in the Milky Way
\citep{brainerd05,azzaro06,yang06, agustsson07}. One recent paper
\citep{sales04} argued that satellites in the 2dfGRS actually
follow Holmberg's suggestion, but this was in error due apparently to
ambiguities in the way position angles are defined in the 2dfGRS
database, and has now been resolved \citep{yang06}.

The anti-Holmberg effect is due to misalignment between the angular
momentum and the triaxial structure of the host halos. A long
literature has now established dark matter halos to be triaxial
objects, with a preference for nearly prolate shapes, and whose
angular momentum is perpendicular to the major axis of the halo
\citep{frenk88,bullock01a,bullock02,jing02,bailin05,hopkins05,bett07}.
If satellites trace the shape of the dark matter halo and if the spin of
the central galaxy disks preserves the direction of the halo angular
momentum, then this would explain the scarcity of satellites near the
rotation axis of the disk.

We examine this in Figure~\ref{figs:Holmb}, where the anisotropy in
the spatial distribution of satellites around primaries is measured by
the distribution of the cosine of the polar angle (measured from the
angular momentum axis of the SUBFIND subhalo centred on the host galaxy).
An isotropic distribution
would be horizontal in this plot, and it is clear that the spatial
distribution of satellites is significantly anisotropic. Simulated
satellites clearly show an ``anti-Holmberg'' effect, aligning
themselves preferentially along the plane perpendicular to the angular
momentum axis of the halo.

The effect is significant but relatively weak; satellites along the
plane outnumber those closer to the rotation axis by roughly
$2$:$1$. The effect depends only very weakly on the luminosity of the
primary, as shown in Figure~\ref{figs:Holmb}, or on the relative
brightness of the satellites. We have also checked that the effect
is essentially independent of the color of the
satellites. Quantitatively, we show this in Table~\ref{tab:costheta},
where we report the {\it average} value of $|cos(\theta)|$ for various
combinations of primary/satellite luminosity/color.

A theoretical study of these alignment issues based on an
N-body/semi-analytic model similar to but substantially smaller than
the Millennium Simulation has recently been carried out by \citet{kang07}.
These authors did not require their primary systems to be
isolated, nor did they require a substantial magnitude difference
between host galaxy and satellites.  As a result it is difficult to
compare their results quantitatively with our own.  Nevertheless,
there are a number of results in common between our two studies.  They
found that the observed alignments between satellite and central
galaxies were best explained by assuming the minor central galaxy to
align with the spin of its host subhalo (as assumed here) rather than
assuming the major axis of the galaxy to align with that of its
subhalo, and they found alignments of similar strength to those we show
here. Their model gave dependences of the strength of the alignment
signal both on the colour of the host galaxy and on the colour of the
satellites. Such trends are weak or absent for the MS samples we study
here. This difference is most likely due to the much wider range of
systems included in the Kang et al study.

The weak dependence we find of the anisotropy on galaxy properties such as
luminosity and color seems at odds with a number of observational
studies. Some of them suggest that only satellites of red primaries
are anisotropically distributed (Azzaro et al. 2006, Yang et al. 2006,
Agustsson \& Brainerd 2007), and that the effect is enhanced when
considering red companions to red primaries (Yang et al 2006). We
note, however, that this result may be affected by the large scatter
in the mass-luminosity-color relation discussed in
\S~\ref{ssec:lmc}. Satellites are typically searched within a fixed
radius (of order $500$ kpc), a choice that would lead to the inclusion
of a larger fraction of interlopers around blue primaries, which tend
to inhabit halos of lower mass. This may dilute the anisotropy in blue
subsamples, explaining the observational results.

A similar comment applies to other potential correlations reported in
the literature, such as the trend for spatial anisotropies to decline
at large radius (Brainerd 2005; Augustsson \& Brainerd 2007) or to
increase with halo mass (Yang et al 2006). As this paper was nearing
submission, \citet{bailin07} argued in a recent preprint that the
alignment of satellites may significantly depend on the isolation
criteria applied. These authors find that satellites of ``truly''
isolated SDSS primaries {\it do} show a polar excess (the original
``Holmberg effect''). We have checked for this in our catalogue, but
find little dependence of our conclusions on $L_{\rm host}/L_{\rm
sys}$. A detailed assessment of the correlations claimed by Bailin et
al (2007) in light of our results is beyond the scope of this paper,
but it is clearly an issue of interest to which we plan to return in
future work.

\begin{center}
\begin{figure}
\includegraphics[width=84mm]{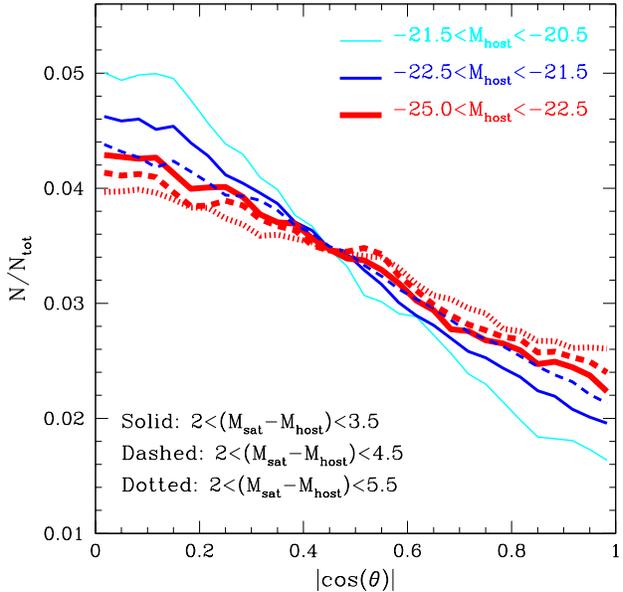}
\caption{Spatial anisotropy of the satellite spatial distribution, as
measured by the distribution of the cosine of the polar angle,
$|cos(\theta)|$, measured from the rotation axis of the host
halo. Different colors correspond to different primary luminosity, as
labelled. Different line types correspond to varying the magnitude
range used to select satellites, also as labelled (see also
Figure~\ref{figs:dprofl}). An isotropic distribution would be
horizontal in this plot. Satellites show a well-defined
``anti-Holmberg'' effect; i.e., they tend to populate preferentially
the plane {\it perpendicular} to the angular momentum axis of the host
halo. }
\label{figs:Holmb}
\end{figure}
\end{center}


\section{Summary and Conclusions}
\label{sec:conc}

We analyse a large sample of isolated galaxies and their satellites
selected from the semianalytic galaxy catalogue constructed by Croton
et al (2006)  from the {\it Millennium Simulation} ({\it MS}). The large
number of galaxies in the catalogue, together with the large volume
surveyed by the {\it MS}, allow us to characterize in detail the 3D
dynamical properties of the satellite population of bright isolated
galaxies. Our isolation criterion stipulates that all galaxies within
$1\, h^{-1}$ Mpc should be at least 2 mag fainter than the
primary. This criterion typically selects galaxies in sparse
environments, but also picks systems of galaxies with peculiar
luminosity gaps between the two brightest galaxies, such as ``fossil
groups''.

Our main conclusions may be summarized as follows:

\begin{itemize}

\item The relation between the halo mass of isolated galaxies and
their luminosity shows very large scatter (halo masses span over a
decade in mass at given luminosity), compromising the ability of
studies that rely on stacking satellites of galaxies of similar
luminosity to probe their dark matter halos. Selecting primaries by
color in order to eliminate the reddest and bluest primaries helps to
tighten the mass-luminosity relation, but still a number of
``underluminous'' central galaxies of massive groups remain in the
sample. These may, however, be excluded by surveying the environment
of the primary and rejecting those in regions of anomalously high
richness.

\item One corollary of the above conclusion is that the relation
between primary luminosity and satellite velocity dispersion is rather
sensitive to the primary selection criteria. Stacking all satellites
of all primaries leads to an $L$-$\sigma$ relation that is poorly
approximated by a power law, and much shallower that the $L\propto
\sigma^3$ scaling expected for hierarchical models. Only after weeding
out massive halos with underluminous central galaxies do we recover
the expected $L \propto \sigma^3$ scaling.

\item Since our isolation criterion readily selects the central
galaxies of ``fossil'' groups, our analysis may be used to predict the
abundance of groups with unusual gaps in the luminosity of the two
brightest galaxies. We find that about $8$ to $10\%$ of halos
exceeding $10^{13} \, M_\odot/h$ would qualify as ``fossil'' systems,
a result that seems consistent with the (so far rather uncertain)
observational constraints. We examine recent claims that the
luminosity function of ``fossil'' groups may be difficult to reproduce
in the $\Lambda$CDM cosmogony but find no obvious discrepancy with
observational constraints for the three fossil group luminosity
functions in the literature. Further data are needed to settle this
issue.

\item The density profile of satellites around primaries may be well
approximated by an NFW profile that is slightly less concentrated than
the average dark matter profile. This conclusion is sensitive to the
color of the satellites; red satellites are significantly {\it more}
concentrated than the dark matter; the opposite is true for {\it blue
} satellites. We also find evidence for luminosity segregation in the
satellite population; i.e., a weak tendency for satellites near the
primary to be brighter than those further away.

\item The velocity distribution of satellites is, like the dark
matter, dominated by radial motions within the virial radius. The
anisotropy is maximal at intermediate radii, becoming gradually more
isotropic near the virial radius. This is a result of the
radially-increasing contribution of satellites on their first infall
onto the primary, a population of objects with rather small dispersion
in radial velocity whose contribution raises the importance of
tangential motions in the outskirts of the host halo.

\item Satellites are distributed anisotropically around primaries,
with a well-defined but relatively weak preference for the plane
perpendicular to the angular momentum of the halo (an
``anti-Holmberg'' effect). This is consistent with the latest
observational studies, and is a direct result of misalignment between
the angular momentum axis and the triaxial structure of dark matter
halos.

\end{itemize}

The characterization of the satellite population of isolated, bright
galaxies we present here has several goals: (i) to guide the
compilation of primary/satellite systems that minimize the presence of
interlopers, (ii) to facilitate the interpretation of observational
results, and (iii) to provide predictions that may be used to validate
the semianalytic model of galaxy formation applied to the {\it
Millennum Simulation}, as well as, more generally, the hierarchical
nature of galaxy assembly. Many of our results are amenable to direct
confrontation with observation, and it is to be hoped that such a
comparison will provide a number of insights into galaxy formation
physics, and perhaps even some challenges to the $\Lambda$CDM
paradigm.

\begin{table*}
\caption{Abundance of {\it primary} galaxies in the {\it Millennium
Simulation} as a function of halo mass. Columns 2 and 3 indicate the
median and the rms dispersion of the luminosity that correspond to
primaries residing in halos within each mass bin. Columns 4 and 5 show
the density ($n_{\rm prim}$) and the fraction ($f_{\rm prim}$) of
halos in the Millennium Simulation that host a {\it primary} galaxy in
our catalogue respectively. The last three columns list the
contribution to $f_{\rm prim}$ in each mass bin after applying the
color cuts shown in Figure~\ref{figs:lmc}.}
\begin{tabular}{|c|c|c|c|c|c|c|c|}
\hline
{$M_{200}$} & ${\log(L_r/L_\odot)}$ & $\sigma(\log(L_r/L_\odot))$ & $n_{\rm prim}$  
& $f_{\rm prim}$ & $f_{g-r}$ (\%) & $f_{g-r}$ (\%) & $f_{g-r}$ (\%) \\
$[M_\odot/h]$ & [median] & & $[h^3$Mpc$^{-3}]$& \% & $< 0.65$ & $<0.95 \ \& >0.65$ & $>0.95$ \\
\hline
$10^{11}-10^{12}$ & 10.17 & 0.15 & $15.04 \times 10^{-4}$& 24.7 & 1.6 & 21.2 & 1.9 \\
$10^{12}-10^{13}$ & 10.51 & 0.18 & $5.50 \times 10^{-4}$& 72.1 & 0.6 & 37.0 & 34.5 \\
$10^{13}-10^{14}$ & 10.85 & 0.18 & $8.37 \times 10^{-6}$& 10.1 & 0.1 & 0.3 & 9.7 \\
$10^{14}-10^{15}$ & 11.34 & 0.16 & $0.34 \times 10^{-6}$& 8.3 & 0.2 & 0.6 & 7.5 \\
\hline
\end{tabular}
\label{tab:mfrac}
\end{table*}
%
\begin{table*}
\caption{Abundance of {\it primary} galaxies in the {\it Millennium
Simulation} as a function of $r$-band luminosity. Columns 2 and 3
indicate the median and the rms dispersion of the halo mass where
primaries of a given luminosity reside. Columns 4 and 5 show the
density ($n_{\rm prim}$) and the fraction ($f_{\rm prim}$) of galaxies
in the Millennium Simulation that are classified as {\it primaries} in
our catalogue respectively. The last three columns list the
contribution to each luminosity bin after applying the color cuts
shown in Figure~\ref{figs:lmc}.  }
\begin{tabular}{|c|c|c|c|c|c|c|c|}
\hline
{$\log{L_r}/L_\odot$} & ${\log(M_{200}/M_\odot)}$ & $\sigma(\log({M_{200}/M_\odot}))$ & $n_{\rm prim}$ 
& $f_{\rm prim}$ & $f_{g-r}$ (\%) & $f_{g-r}$ (\%) & $f_{g-r}$ (\%) \\
 & [median] & & $[h^3$Mpc$^{-3}]$ & \% & $< 0.65$ & $<0.95 \ \& >0.65$ & $>0.95$ \\
\hline
$10.0-10.2$& 11.75 & 0.17 & $9.00 \times 10^{-4}$ & 18.8 & 1.3  & 15.03 & 2.5  \\
$10.2-10.4$& 11.95 & 0.20 & $6.22 \times 10{-4}$  & 22.1 & 0.7  & 16.5  & 4.9 \\
$10.4-10.6$& 12.19 & 0.23 & $3.82 \times 10^{-4}$ & 24.8  & 0.5  & 16.9  & 6.1 \\
$10.6-10.8$& 12.44 & 0.25 & $1.39 \times 10^{-4}$ & 18.4  & 0.9  & 15.1  & 7.3  \\
$10.8-11.0$& 12.64 & 0.37 & $2.03 \times 10^{-5}$ & 25.3  & 5.0  & 11.7  & 7.8 \\
$>11$      & 13.34 & 0.76 & $3.70 \times 10^{-6}$ & 25.1  & 15.6  & 1.92  & 7.8  \\
\hline
\end{tabular}
\label{tab:lfrac}
\end{table*}
%

\begin{table*}
\caption{Moments of the satellite velocity distributions shown in
Figures~\ref{figs:rvdist} and ~\ref{figs:vtandist} in several distance
bins. The last column show $f_{\rm inf}$, the fraction of satellites in
each radial shell that are in their first approach to the host.}
\begin{tabular}{|c|c|c|c|c|c|c|}
\hline
Component & $r/r_{200}$ & mean & sigma ($\sigma$) & skewness ($\xi$) & kurtosis ($\kappa$) & $f_{\rm inf}$\\
\hline
$V_r$ & 0.0-0.5 & -0.014 & 0.96 & 0.025 & -0.68 & --\\
& 0.5-1.0 & -0.072 & 0.70 & -0.002 & -0.45 & --\\
& 1.0-1.5 & -0.133 & 0.54 & 0.014 & 0.23 & 15\%\\
& 1.5-2.0 & -0.187 & 0.51 & 0.257 & 2.39 & 25\% \\
& 2.0-2.5 & -0.208 & 0.52 & 0.687 & 5.97 & 36\% \\
& 2.5-3.0 & -0.179 & 0.55 & 0.644 & 8.22 & 40\% \\
\hline
$V_\theta$ & 0.0-0.5 & 0.0 & 0.45 & 0.004 & 0.99&--\\
& 0.5-1.0 & 0.0 & 0.39 & 0.058 & 1.44&--\\
& 1.0-1.5 & 0.0 & 0.49 & -0.087 & 2.30& 15\%\\
& 1.5-2.0 & 0.0 &  0.51 & 0.143 & 8.82& 25\%\\
& 2.0-2.5 & 0.0 & 0.55 & 0.002 & 16.58& 36\%\\
& 2.5-3.0 & 0.0 & 0.60 & 0.240 & 20.35& 40\%\\
\hline
$V_\phi$ & 0.0-0.5 & 0.20 & 0.x55 & -0.020 & 0.48&--\\
& 0.5-1.0 & 0.25 & 0.50 & -0.240 & 1.15 & --\\
& 1.0-1.5 & 0.22 & 0.51 & -0.456 & 2.23 &  15\%\\
& 1.5-2.0 & 0.15 & 0.52 & -0.387 & 4.00 &  25\%\\
& 2.0-2.5 & 0.10 & 0.57 & -0.785 & 18.23 & 36\%\\
& 2.5-3.0 & 0.05 & 0.60 & -0.015 & 15.00 & 40\%\\
\hline
\end{tabular}
\label{tab:Gaussfits}
\end{table*}
\begin{table*}
\caption{Average $|cos(\theta)|$ (where $\theta$ is the polar angle
measured from the halo angular momentum axis) of all satellites for
different bins in host luminosity and color.  The last three columns
show the values corresponding to three equal-number subsamples: blue
satellites ($(g-r)_{\rm sat} < 0.95$), intermediate color satellites
($0.95<(g-r)_{\rm sat}< 1.01$) and red satellites ($(g-r)_{\rm sat}
>1.01$).}
\begin{tabular}{|c|c|c|c|c|}
\hline
$M_{\rm host}$ & $\langle{|cos(\theta)|}\rangle$ & $\langle{|cos(\theta)|}\rangle$ & $\langle{|cos(\theta)|}\rangle$ & $\langle{|cos(\theta)|}\rangle$\\
& all & $(g-r)_{\rm sat}<0.95$ & $0.95<(g-r)_{\rm sat}<1.01$ & $(g-r)_{\rm sat}>1.01$ \\ 
\hline
$-21.5<M_{\rm host}<-20.5$ & 0.410 & 0.404 & 0.408 & 0.422\\
$-22.5<M_{\rm host}<-21.5$ & 0.410 & 0.447 & 0.441 & 0.439\\
$M_{\rm host}<-22.5$ & 0.410 & 0.476 & 0.463 & 0.452\\
\hline
$(g-r)_{\rm host}$ & & & & \\
\hline
$(g-r)_{\rm host}\leq0.65$ & 0.426 & 0.437 & 0.421 & 0.417\\
$0.65<(g-r)_{\rm host}\leq0.95$ & 0.428 & 0.447 & 0.430 & 0.436\\
$(g-r)_{\rm host}>0.95$ & 0.426 & 0.497 & 0.458 & 0.450\\
\hline
\end{tabular}
\label{tab:costheta}
\end{table*}
%

\section*{Acknowledgments}

LVS would like to thank Dr. Mario Abadi for many helpful suggestions
and discussions. This work was partially supported by the Latin
American European Network on Astrophysics and Cosmology of the
European Union's ALFA Programme and the Consejo Nacional de
Investigaciones Cient\'{\i}ficas y T\'ecnicas (CONICET),
Argentina. LVS is grateful for the hospitality of the Max-Planck
Institute for Astrophysics in Garching, Germany, where much of the
work reported here was carried out. Data for the galaxy formation
model on which this study is based are available at 
{\small \tt http://www.mpa-garching.mpg.de/galform/agnpaper/}
Data for other galaxy formation models and
for the halo/subhalo populations of the {\it Millennium Simulation}
are available for all redshifts at {\small \tt
http://www.mpa-garching.mpg.de/millennium}.
We acknowledge Eduardo Cypriano for sending us electronic
data for the fossil groups RX1416 and RX1552. We also thank
the anonymous referee for useful suggestions and comments
that helped to improve the previous version of the paper.

\bibliography{references}

\end{document}